\documentclass[journal=jpcb,manuscript=article,layout=twocolumn]{achemso}

\usepackage[version=3]{mhchem} 
\usepackage{graphicx}
\usepackage{dcolumn}
\usepackage{bm}
\usepackage[utf8]{inputenc}
\usepackage[T1]{fontenc}
\usepackage{mathptmx}
\usepackage{amsmath}
\usepackage{amssymb}
\usepackage{etoolbox}
\usepackage{siunitx}
\usepackage{subcaption}

\newcommand*{\down}{\textsubscript}

\newcommand*{\diff}{\mathop{}\!\mathrm{d}}

\renewcommand{\vec}[1]{{\boldsymbol{#1}}}
\DeclareSymbolFont{rmlargesymbols}{OMX}{mdbch}{m}{n}
\DeclareMathSymbol{\rmintop}{\mathop}{rmlargesymbols}{82}

\setkeys{acs}{maxauthors=10, etalmode=truncate}

\newcommand{\numertraj}  {S1}
\newcommand{\SIprocrustessimilaritycomparison}{S1}
\newcommand{\SIconstraintsimilaritydistributioncomp}  {S2}
\newcommand{\SIrestraintsimilaritydistributioncomp}  {S3}
\newcommand{\SIStbconstraintsetAmiscoremat}  {S4}

\newcommand{\SIStbconstraintsetCmiscoremat}  {S6}

\newcommand{\SIAAPanelQ}  {S10}
\newcommand{\SIAAgeometricdG}{S11}
\newcommand{\SIAAfigoneworkQQ}{S12}
\newcommand{\SIAAfigtwoworkQQ}{S13}

\author{Miriam J\"ager}
\author{Steffen Wolf}
\affiliation{ 
Biomolecular Dynamics, Institute of Physics, University of Freiburg, 79104 Freiburg, Germany
}%
 \email{steffen.wolf@physik.uni-freiburg.de.}
\phone{+49 (0)761 203 5913}

\title{More sophisticated is not always better:\\comparison of similarity measures for unsupervised learning of pathways in biomolecular simulations}

\abbreviations{MD, TMD, dcTMD, SMD, PCA, CV, DTW}
\keywords{biased molecular dynamics simulations, protein-ligand unbinding, pathways}

\begin{document}


\begin{abstract}
Finding process pathways in molecular simulations such as the unbinding paths of small molecule ligands from their binding sites at protein targets in a set of trajectories via unsupervised learning approaches requires the definition of a suitable similarity measure between trajectories. We here evaluate the performance of four such measures with varying degree of sophistication, i.e., Euclidean and Wasserstein distances, Procrustes analysis and dynamical time warping, when analyzing trajectory data from two different biased simulation driving protocols in the form of constant velocity constraint targeted MD and steered MD.
In a streptavidin-biotin benchmark system with known ground truth clusters, Wasserstein distances yielded the best clustering performance, closely followed by Euclidean distances, both being the most computationally efficient similarity measures. In a more complex A\textsubscript{2a} receptor-inhibitor system, however, the simplest measure, i.e., Euclidean distances, was sufficient to reveal meaningful and interpretable clusters.
\end{abstract}

\section{Introduction}

The prediction of formation and dissociation pathways of biomolecular complexes via molecular dynamics (MD) simulations is an active research field in computational biophysics. Understanding the associated complex formation and decay mechanisms as well as process rates holds the potential to enable their targeted manipulation.\cite{Bruce18,Bernetti19,Decherchi20,Ahmad22,Sohraby22,Conflitti23,Wang23a,Wolf23a,Kordylewski2025} Prominent examples include tailoring the binding and unbinding kinetics of drugs to improve their efficacy,\cite{Copeland2006,Swinney06,Schuetz17} while selectively blocking oxygen diffusion channels in hydrogenases may improve the protein's resistance against this aggressive element,\cite{Sohraby24} allowing for the biotechnical generation of ''green'' hydrogen. Lastly, oncogenic mutants of kinases causing drug resistance may not change the affinity of a compound, but accelerate its unbinding.\cite{Lyczek21}

Predicting such processes with MD simulations comes with two main challenges: first, due to the inherent timescales of unbinding on the order of seconds to hours in the case of protein-ligand complexes, unbiased brute-force MD cannot reveal these dynamics. Therefore, biased MD approaches are the main access to sufficiently large samples of unbinding events.\cite{Henin2022,Wolf23a} Second, describing binding and unbinding events requires identifying the pathways a ligand takes to traverse to and from its binding site, respectively, and determining the path collective variable (CV) space in which these pathways are found. In the case of biased MD simulations, besides approaches that learn path CVs with the help of artificial intelligence on-the-fly,\cite{Rohrdanz2013,Wang19,Rydzewski19,Jung23,Ray23b,Qiu23,Falkner2024,Megias2025} pathways are usually identified a posteriori in the form of classes of trajectories that are ''similar'' based on a criterion such as Hausdorff and Fr\'echet metrics,\cite{Seyler15} interaction and contact fingerprints,\cite{Kokh20,Bianciotto21} ligand RMSD,\cite{Bray2022b} protein-ligand contact principal components (conPCA)\cite{Ernst2015,Wolf20,Taenzel24} or dynamical time warping (DTW).\cite{Ray24} So far, no comparison of the performance of different similarity criteria using a suitable benchmark system has been conducted. Furthermore, the different criteria can be expected to vary in performance based on the biasing protocol used. For example, while a velocity constraint-bias used in dissipation-corrected targeted MD\cite{Wolf18} has been shown to perform well with a conPCA-derived distance,\cite{Taenzel24} DTW appears to work well in combination with infrequent Metadynamics.\cite{Tiwary15,Ray24} 

In this work, we evaluate the performance of different similarity measures for the unsupervised classification of sets of protein-ligand unbinding trajectories into pathways based on ligand-protein contact distances as input features. As a set of measures with varying complexity and numerical demands, we utilize Euclidean conPCA distances,\cite{Taenzel24} Wasserstein and Procrustes metrics\cite{Yuan17}, as well as DTW.\cite{Ray24} 
We apply all measures to a set of trajectories from enforced unbinding simulations of the streptavidin-biotin complex\cite{Cai2023} with known ground truth of unbinding pathways. As we herein have defined unbinding pathways beforehand in the pulling protocol, this dataset constitutes an excellent benchmark for pathway separation approaches.
Employing both constant velocity constraint-based targeted MD\cite{Schlitter1994} and harmonic restraint-based steered MD\cite{Isralewitz01} simulations, we investigate the applicability of the similarity measures to different biasing protocols\cite{Henin2022} with varying coupling between bias and system. 
To rule out observational bias from only investigating a single protein-ligand complex, we then extend the investigation to simulations of 
an A\down{2a} receptor-inhibitor complex.\cite{Segala2016} 
As we identified ligand-membrane peptide contacts to discriminate between possible pathways in this system in an earlier study,\cite{Taenzel24} we hereby check if the investigated distance measures are suitable to reveal mechanistically relevant pathways.
We furthermore compare the potential of mean force along the identified pathways to a reference free energy obtained by dissipation-corrected targeted MD.\cite{Wolf18,Taenzel24}

\section{Theory and Methods}

\subsection{Targeted and Steered Molecular Dynamics}
In targeted molecular dynamics\cite{Schlitter1994} (TMD) and steered molecular dynamics\cite{Izrailev99} (SMD) simulations, the goal is to accelerate otherwise slow processes, such as ligand unbinding, by applying an external bias. This manipulation enforces rare events, thereby uncovering diverse binding or unbinding routes that would be difficult to observe in equilibrium simulations.
In TMD, the system (e.g., a ligand) is forced along a (here one-dimensional) biasing coordinate $x$ with a constant velocity $v$. This biasing is implemented via a time-dependent distance constraint
$\Phi(t) = x (t) - (x_0 +v t) \overset{!}{=} 0$, where the constraint enters the MD integrator as a force $f_c = \lambda (\diff{\Phi}/\diff{x})$ with a Lagrange multiplier $\lambda$. 
SMD also involves pulling the system along a chosen biasing coordinate, but is implemented via a harmonic spring force $f_\text{ext}(t) = -k (x(t)-x_0-v t)$
with the spring constant $k$. In contrast to the numerically exact linear connection between $x$ and $t$ in TMD, SMD allows for fluctuations of $x(t)$ around $x_0+vt$, introducing additional difficulties for a posteriori path finding approaches. Similar fluctuations can be found in other biasing protocols such as infrequent Metadynamics,\cite{Tiwary15} ligand Gaussian-accelerated MD\cite{Miao20} or tauRAMD,\cite{Kokh18} allowing for an extrapolation of our results to these approaches.

\subsection{Pathway separation of trajectories}

\subsubsection{Input Features}

Following the approach introduced in Ref.~\citenum{Taenzel24}, ligand–protein unbinding trajectories are represented using internal coordinates as input features, here ligand–protein contact distances.\cite{Ernst2015}
A residue is considered a contact if its C\textsubscript{$\alpha$} atom is within \SI{0.45}{nm} of any ligand atom at any point during the trajectories. The contact distances are the minimal distances between any ligand and protein amino acid heavy atoms.
Thus, each trajectory is encoded as a matrix of size $K \times M$, where rows $K$ represent time steps and columns $M$ correspond to distinct contact distances. Given a set of $ N $ trajectories, we define the dataset as
\begin{align}\label{eq:contactdistancedataset}
    \mathcal{P} = \left\{ P^{(i)} \in \mathbb{R}^{K \times M} \mid i = 1, \dots, N \right\},
\end{align}
where each matrix $ P^{(i)} = \left[ p_{t,m}^{(i)} \right] $ corresponds to the $i$-th trajectory, with index $ t \in \{1, \dots, K\} $ denoting time-steps and $ m \in \{1, \dots, M\} $ denoting contacts. Alternatively, each trajectory $i$ can also be represented as time dependent vector $\mathbf{p^{(i)}}(t) = (p_{1}^{(i)}(t), p_{2}^{(i)}(t), \dots, p_{M}^{(i)}(t))^\top$.

\subsubsection{Trajectory Preprocessing}
    For an improved data analysis and noise filtering, we tested the influence of different preprocessing routines for the contact distances on the clustering outcomes:
    \paragraph{Smoothing}
    To reduce high-frequency noise in the raw contact distance time series, a temporal Gaussian filter can be applied to each trajectory.\cite{Nagel23} For each fixed contact distance $m$, the smoothed value at time step $t$ is computed as
    \begin{align}\label{eq:gaussfilter}
    \tilde{p}_{t,m}^{(i)} = \sum_{\nu=-\sigma}^{\sigma} w_\nu \cdot p_{t+\nu,m}^{(i)}
    \end{align}
    where $w_\nu$ are weights from a normalized Gaussian kernel and $\sigma$ defines the filter window size.

    \paragraph{Time-resolved and global normalization}
    To counteract the influence of large inter-trajectory distances resulting from free diffusion in the unbound state, each contact distance $m$ can be scaled with the mean over $N$ trajectories evaluated at each step $t$.\cite{Bray2022b, Taenzel24}
    We scale each contact distance $p_{t,m}^{(i)}$ by the mean of its corresponding matrix element across all trajectories at each time point $t$:
    \begin{align}\label{eq:normT}
        \mu_{t,m} = \left\langle  p_{t,m}^{(i)} \right\rangle_i,
        \quad\quad \hat{p}_{t,m}^{(i)} = \frac{p_{t,m}^{(i)}}{\mu_{t,m}}
    \end{align}
    where $\langle \cdot \rangle_i$ denotes the average over all $N$ trajectories.
    In addition to the per-element normalization described in Eq.~\eqref{eq:normT}, which normalizes each contact distance independently, we can also apply a global normalization.
    To bring all input features to a common scale, we apply the following normalization to the contact distances
    \begin{align}
    \mu_m = \left\langle p_{t,m}^{(i)} \right\rangle_{i,t}, \quad 
    \label{eq:timedistmean}
    \sigma_m = \sqrt{ \frac{1}{NK} \sum_{i=1}^{N} \sum_{t=1}^{K}\left( p_{t,m}^{(i)} - \mu_m \right)^2 }
    \end{align}
     where $\langle \cdot \rangle_{i,t}$ denotes the average over all $N$ trajectories and all time-steps.
    The standardized entry is given by
    \begin{align}\label{eq:global_norm}
    \hat{\hat{p}}_{k,m}^{(i)} = \frac{p_{k,m}^{(i)} - \mu_m}{\sigma_m},
    \end{align}
    which ensures that each contact distance has zero mean and unit variance.

    \paragraph{Principal Component Analysis}
    To reduce the dimensionality of the contact distance data and extract putative path CVs, we carry out a Principal Component Analysis (PCA) following the approach introduced in Refs.~\citenum{Ernst2015, Post2019, Taenzel24}.
    The covariance matrix of the contact distances across all trajectories and time steps is defined as
    \begin{align}
    \sigma_{m,m'} = \frac{1}{NK} \sum_{i=1}^{N} \sum_{t=1}^{K} \left( p_{t,m}^{(i)} - \mu_m \right) \left( p_{t,m'}^{(i)} - \mu_{m'} \right),
    \end{align}
    where $\mu_m$ is given by Eq.~\eqref{eq:timedistmean}.
    Diagonalization of the covariance matrix $(\sigma_{m,m'})$ produces a set of $k$ eigenvectors $\{\mathbf{e}_k\}$, arranged according to the magnitude of their eigenvalues $\{\lambda_k\}$ in decreasing order. To obtain the principal components $\mathrm{PC}^{(i)}_{k}(t)$, the contact distances $\mathbf{p}^{(i)}(t)$ are projected onto the eigenvectors via
    $\mathrm{PC}^{(i)}_{k}(t) = \mathbf{e}_k \cdot \mathbf{p}^{(i)}(t)$.
    Since a PCA is a unitary transformation, the projection preserves lengths as well as dimensions. Additionally, the composition of the eigenvectors $\{\mathbf{e}_k\}$ reveals the relative contributions of individual contacts. This allows to identify key input features and may yield insights into the microscopic discriminants of the unbinding process.\cite{Wolf2020,Wolf2023}

\subsection{Similarity Measures}

Comparing molecular trajectories to identify clusters that follow the same pathway requires a well-defined notion of similarity or dissimilarity between time series representing structural observables.\cite{Taenzel24} This section introduces several distance-based approaches for this purpose.

\paragraph{Euclidean Distances} are the most straightforward distance measure. To obtain a scalar value for each trajectory pair we used the average over all time steps. Given two trajectories $ P^{(i)}, P^{(j)} \in \mathbb{R}^{K \times M} $ with matrix elements $p_{t,m}$ and $q_{t,m}$, respectively,
the mean Euclidean distance is defined as
\begin{align}\label{eq:euclideandistance}
    d_{\text{E}}(P^{(i)}, P^{(j)}) = \left\langle \sqrt{\sum_{m=1}^{M} \left( p_{t,m} - q_{t,m} \right)^2} \right\rangle_{t}
\end{align}
with $\langle \cdot \rangle_t$ denoting the average over all time steps.
This distance is computationally efficient, scaling linearly with the number of time steps and dimensions, and is easy to interpret geometrically. However, its applicability is limited to trajectories of equal length with well-aligned time points. This requirement is satisfied for constrained pulling simulations that start from the same center-of-mass distance $d_{\rm com}$, but may be violated in other settings such as restraint-based pulling or unbiased simulations. In such cases, more flexible alignment schemes are needed.

\paragraph{Dynamic Time Warping (DTW)}\cite{Marteau2008}, unlike the Euclidean distance, allows non-linear alignments in time even for trajectories with unequal lengths, making it particularly suitable for comparing trajectories that evolve at different rates but share a common progression of events.
It was originally introduced in speech recognition\cite{Sakoe1978} and has been widely adopted for time series analysis.
The standard DTW algorithm is used to distinguish one-dimensional time series. It can be generalized to a multi-dimensional case by treating each feature separately (independent DTW). In MD context, however, inter feature correlations  (e.g., between distances to neighboring residues) occur often. 
Dependent DTW (DTW\textsubscript{D}) addresses this by computing vector norms at each alignment step, better capturing the multivariate trajectory geometry.\cite{ShokoohiYekta2017} DTW has been used in MD to distinguish between transition paths\cite{Ray24} and conformational states.\cite{Schuhmann2023}
Given trajectories $ P^{(i)} \in \mathbb{R}^{K \times M}$, $P^{(j)} \in \mathbb{R}^{K' \times M} $ with entries $p_{t,m}$ and $q_{t',m}$ respectively, the dependent DTW (DTW\textsubscript{D})\cite{ShokoohiYekta2017} constructs a matrix $ D \in \mathbb{R}^{K \times K'}$:
\begin{align*}
D(t,t') = &d_e^2(\vec{p}(t), \vec{q}(t')) \\
            &+ \min \left\{ D(t-1,t'-1), D(t-1,t'), D(t,t'-1) \right\}
\end{align*}
with the euclidean distance $d_e^2(p_t,q_{t'})=\sum_m(p_{t,m}-q_{t',m})^2$,
A warping path $ \pi  = \{ (t_1, t'_1), \ldots, (t_T, t'_T) \}$ is defined as a sequence of matrix indices that represents an alignment between the trajectories. The path must satisfy boundary conditions on the time steps $ (t_1, t'_1) = (1, 1) $, $ (t_T, t'_T) = (K, K') $, monotonicity constraints $ t_{l+1} \geq t_l $, $ t'_{l+1} \geq t'_l $ and the steps in $\pi $ are restricted to a adjacent cells. 
The optimal alignment path is obtained by minimizing the cumulative cost for stepping through $D(t,t')$ recursively. Thus, the final DTW distance is
\begin{align}\label{eq:DTW}
d_{\text{DTW}}(P^{(i)}, P^{(j)}) = \min_{\pi \in \Pi} \sum_{D(t,t') \in \pi} D(t,t')
\end{align}
where $\Pi$ is the set of all possible warping paths on $D$.

DTW is not a true metric as it violates the triangle inequality, but it is robust to temporal misalignments and less sensitive to local noise or outliers. 

\paragraph{Procrustes Analysis}\cite{Gower1975}
is a method used to compare two point clouds or matrices by finding the optimal linear transformation (translation, scaling and rotation) that best superimposes one data set onto the other.
Given trajectories represented as matrices $ P^{(i)}, P^{(j)} \in \mathbb{R}^{K \times M} $, the Procrustes distance is defined as:
\begin{align}\label{eq:procrustes}
    d_{\text{P}}(P^{(i)}, P^{(j)}) = \min_{s, \mathbf{R}, \mathbf{t}} \left\| s P^{(i)} \mathbf{R} + \mathbf{1}_K \mathbf{t}^\top - P^{(j)} \right\|_F
\end{align}
where $ s \in \mathbb{R} $ is a scalar scaling factor, $ \mathbf{R} \in \mathbb{R}^{M \times M} $ is an orthogonal rotation matrix, $ \mathbf{t} \in \mathbb{R}^{M} $ is a translation vector, $ \mathbf{1}_K \in \mathbb{R}^K $ is a vector of ones, and $ \left\| \cdot \right\|_F $ denotes the Frobenius norm. This distance captures differences in shape rather than absolute position, making it suitable for cases where trajectories share geometric structure. 

\paragraph{1D Wasserstein Distance} (Earth Mover’s Distance)
\cite{Villani2009}
quantifies the cost of transforming one probability distribution into another.  It has been used to compare molecular conformations\cite{GonzalezDelgado2023} and protein-ligand interactions.\cite{Mustali2023} It does not require time alignment and treats each trajectory as a distribution over structural configurations.
Here, we treat the contact distances $p_{t,m}=\{p_{1,m}, p_{2,m}, \dots, p_{K,m}\}$ as independent identically distributed (i.e., the associated weights $w_t$ to each time step $t$ are $w=1/K$) random numbers and determine their empirical cumulative distribution function $F_{p_m}$.
The 1D-Wasserstein distance between trajectories $ P^{(i)} $ and $ P^{(j)} $ with elements $p_{t,m}$ and $q_{t,m}$ respectively, for a given contact distance $m$ is:
\[
W_m(P^{(i)}, P^{(j)}) = \int_{0}^{1} | F^{-1}_{p_m}(u) - F^{-1}_{q_m}(u)| \, du
\]
where $F^{-1}_{p_m}(u)$ is the inverse of the empirical cumulative distribution function (CDF) of the contact distance $p_m$ at quantile $u$. The Wasserstein distance is a true metric, satisfying the triangle inequality. 
Summing across dimensions yields the distance for the whole trajectory
\begin{align}
\label{eq:Wasserstein}
    d_{\text{W}}(P^{(i)}, P^{(j)}) = \sum_{m=1}^{M} W_m(P^{(i)}, P^{(j)})
\end{align}

\paragraph{Similarity Matrix}
From the pairwise distances between all trajectories, we construct a distance matrix $\mathbf{D}=(d_{i,j})_{i=1,\dots,N;~j=1,\dots,N}$, where each entry $d_{i,j}$ represents the distance between a trajectory pair $P^{(i)}$ and $P^{(j)}$. From the distance matrix we construct a similarity matrix $\mathbf{S}=(s_{i,j})$ on the interval $[0,1]$ via
\begin{align}\label{eq:similarity}
    s_{i,j} = 1 - \frac{d(P^{(i)}, P^{(j)})}{d_{\max}}
\end{align}
where $d_{\max}$ corresponds to the maximum distance in $\mathbf{D}$.

\paragraph{Computational details}
All distance calculations were performed using parallelized code executed on 12 threads of an AMD Ryzen 9 7950X 16-core CPU work station with 32 GB RAM.
Protein renders were generated with \texttt{PyMol}.\cite{PyMOL} Plots were created using \texttt{matplotlib}\cite{matplotlib} and the wrapper \texttt{prettypyplot}.\cite{prettypyplot} Sankey plots were created with an adapted version of \texttt{pysankey}.\cite{pysankey} 
The contact distances were determined from Gromacs trajectories using \texttt{MDAnalysis}.\cite{MDAnalysis2016}
All analyses were performed in Python using \texttt{numpy}.\cite{numpy2020} The distances were calculated using the \texttt{scipy}\cite{SciPy2020} modules
\texttt{stats.wasserstein\_distance} and \texttt{spatial.procrustes}. The PCA was carried out using the \texttt{scikit-learn}\cite{scikit-learn} module \texttt{decomposition.PCA}. The DTW distance was calculated using \texttt{DTAIDistance}.\cite{DTAIDistance} The clustering was performed using \texttt{mosaic.Clustering}.\cite{Diez2022} and clusterings evaluated using \texttt{sklearn.metrics.normalized\_mutual\_info\_score}.

\subsection{Clustering via Leiden community detection}
To cluster trajectories based on their pairwise similarity $s_{ij}$, we use the Leiden community detection algorithm.\cite{Traag2011,Traag2019,Diez2022} This method encodes the similarity matrix $\mathbf{S}$ as a graph, where the nodes represent the trajectories and the edges between nodes are the similarities $s_{ij}$. 
Clustering is performed by maximizing an objective function, for which we employ the Constant Potts Model (CPM)
\begin{equation}\label{eq:CPM}
    \Phi_\mathrm{CPM} = \sum_{c} \left( e_{c} - \gamma \binom{n_{c}}{2} \right).
\end{equation}
Here, $e_{c}$ is the sum of all similarities within a cluster $c$, and $n_{c}$ is the number of trajectories in that cluster. The binomial coefficient $\binom{n_{c}}{2}$ describes a cluster of the same size as $c$ with all pairwise similarities equal to the resolution parameter $\gamma$. Maximizing the above function equals looking for clusters whose summed similarities $e_c$ exceed that lower bound.
The choice of $\gamma$ controls the clustering: 
higher values lead to smaller clusters with a higher intra-cluster similarity, while lower values lead to coarser groupings. Importantly, $\gamma$ does not act as a hard cutoff: if the overall objective benefits from it, some similarities within a cluster can be below $\gamma$.
Applying the Leiden/CPM clustering to a similarity matrix $\mathbf{S}$, we obtain a block matrix (see Fig.~\ref{fig:Stb_Panel1}B) where each block corresponds to a cluster of trajectories that follow the same unbinding pathway. 

\subsubsection{Scoring the clustering results}
In order to evaluate the clustering results we use 
the normalized mutual information (NMI) score.\cite{Vinh2010} NMI quantifies the amount of shared information between two clusterings with a range between 0 (no agreement between clusterings) to 1 (perfect agreement). 
Given a set of N elements with two partitions/clusterings of $C=\{C_1, C_2, ...,C_R\}$ with $R$ clusters and $K=\{K_1, K_2, ..., K_S\}$ with $S$ clusters, where set $C$ is the ground truth cluster assignment, we suppose an element is picked at random. The probability that it falls into cluster $K_i$ is $P_k(i)=|K_i|/N$ with the number of trajectories within the cluster $|K_i|$ and the associated entropy is $H(K)= -\sum_{i=1}^{S} P_K(i) \log(P_K(i))$. Similarly, the probability that an element falls into cluster $C_j$ is $P_C(j)=|C_j|/N$ and the associated entropy is $H(C)= -\sum_{j=1}^{R} P_C(j) \log(P_C(j))$.
The mutual information score between two clusterings $K$ and $C$ is then defined as
\begin{align}\label{eq:mutual_information}
    \text{MI}(K;C) = \sum_{i=1}^{|K|} \sum_{j=1}^{|C|} P_{K,C}(i,j) \log\left(\frac{N|K_i \cap C_j|}{|K_i||C_j|}\right)
\end{align}
where $P_{K,C}(i,j) = |K_i \cap C_j|/N$ is the joint probability of an element being in cluster $K_i$ and $C_j$. The NMI is then calculated as 
\begin{align}\label{eq:normalized_mutual_information}
    \text{NMI}(K;C) = \frac{\text{MI}(K;C)}{[H(K)+H(C)]/2}.
\end{align}

\subsection{Investigated protein–ligand systems and trajectory sets}
We investigate two different biomolecular test systems: First, a set of streptavidin-biotin restraint and constraint pulling trajectories from our previous studies.\cite{Cai2023,Taenzel24}
This trajectory dataset provides a particularly useful benchmark for pathway detection methods because the protein itself undergoes minimal conformational shifts during enforced ligand unbinding, and the ligand is pulled out of the protein along well defined vectors, thus the pathways are well characterized in Cartesian space.

We created three test sets, labeled A, B and C, from the different pulling directions. Set A comprises of trajectories pulled along Cartesian vectors (1,1,0) and (1,-1,0), whereas set B combines (1,0,1) and (1,-1,0). As shown by the density isosurfaces (volume maps) in Fig.~\ref{fig:Stb_Panel1}A, the trajectories in set A are well separated, while those in set B exhibit substantial overlap. This is also evident from the similarity distributions in Fig.~\ref{fig:Stb_Panel1}B, where set B has a higher median similarity. Consequently, the original pulling directions are relatively easy to distinguish in set A, whereas set B presents a more challenging case.
Furthermore, we generated a third data set, set C, by merging all pulling directions from sets A and B: (1,1,0), (1,0,1), and (1,-1,0) to further increase the difficulty of separating each subset.
As stated above, using a harmonic potential-derived force in SMD with 
a force constant of $k$=1000~kJ/(mol~nm) leads to the center of mass distance between the pull groups and thus biotin position to fluctuate around the potential mid-point position $x_0+vt$. In TMD simulations, the center of mass distance grows strictly linear with $t$, which at the same time allows less freedom in dynamics. To explicitly take into account this difference in fluctuations as a challenge for the investigated similarity measures, we wrote out protein-ligand structures from both SMD and TMD simulations with a constant time interval of $\Delta t$=1~ps and 2~ps, respectively, for further contact distance analysis.

In addition to St-b, we employ a membrane-bound ligand–protein system, an inhibitor (ZM241385, in the following abbreviated as ZMA) bound to the A\textsubscript{2a} adenosine receptor.\cite{Segala2016,Taenzel24} In a previous investigation,\cite{Taenzel24} we have examined the free-energy profiles and unbinding routes in considerable detail, making this system a valuable additional test cases for our clustering methodology.

MD simulation details for both systems can be found in Refs.~\citenum{Cai2023,Taenzel24}. The number of  analyzed streptavidin-biotin trajectories is given in the Supplementary Information and Tab.~\numertraj.

\subsection{Dissipation-corrected targeted MD}

The main motivation for our interest in pathway detection is the definition of trajectory clusters following the same pathway for analysis with dissipation-corrected targeted MD (dcTMD):\cite{Wolf2018,Wolf2023} from the required work $W(x)=\int_{x_0}^{x} \mathrm{d}x' \, f_c(x')$ in a set of constant velocity constraint targeted MD simulations, we estimate the free energy $\Delta G(x)$ using the second-order cumulant expansion of the Jarzynski equality\cite{Jarzynski2004,Park2004}
\begin{align} 
    \Delta G(x)\approx\bigl\langle W(x)\bigr\rangle_N - \frac{1}{2\,k_B T}\bigl\langle \delta W(x)^2\bigr\rangle_N. 
\end{align} 
$\langle \cdot \rangle_N$ is the ensemble average over $N$ independent realizations initiated from an equilibrium state, $k_B$ is the Boltzmann constant, and $T$ is the temperature ($k_BT=\beta^{-1}$). If $W(x)$ is normally distributed, this expansion is exact, and one can define the dissipative work as $W_\text{diss}(x) = \frac{1}{2\,k_B T}\,\bigl\langle \delta W(x)^2\bigr\rangle_N$.
Previous work\cite{Wolf2023} demonstrates that if friction depends on the specific route taken in a multidimensional space of path collective variables, the Gaussian work assumption may break down. In protein–ligand systems, this typically manifests in multiple distinct routes or conformational changes that inflate the dissipative work estimate $W_{\rm diss}$.\cite{Wolf2020, Wolf2023, Jaeger2021, Bray2022b}
We classify the pulling trajectories based on their similarity $s$ in path CV space, grouping those with high internal similarity and low cross-group similarity. Each group $k$ is assumed to share a single, well-defined route and thus pathway, resulting in a Gaussian work distribution.\cite{Wolf2023} For each pathway $k$ containing $N_k$ trajectories, the free energy is then computed as
$\Delta G_k (x) = \bigl\langle W(x)\bigr\rangle_{N_k} - \frac{1}{2\,k_B T}\,\bigl\langle \delta W(x)^2\bigr\rangle_{N_k}$ 
yielding pathway-specific $\Delta G_k(x)$.

\section{Results and Discussion}

\begin{figure}[htb!]
    \centering
    \includegraphics[width=0.95\linewidth]{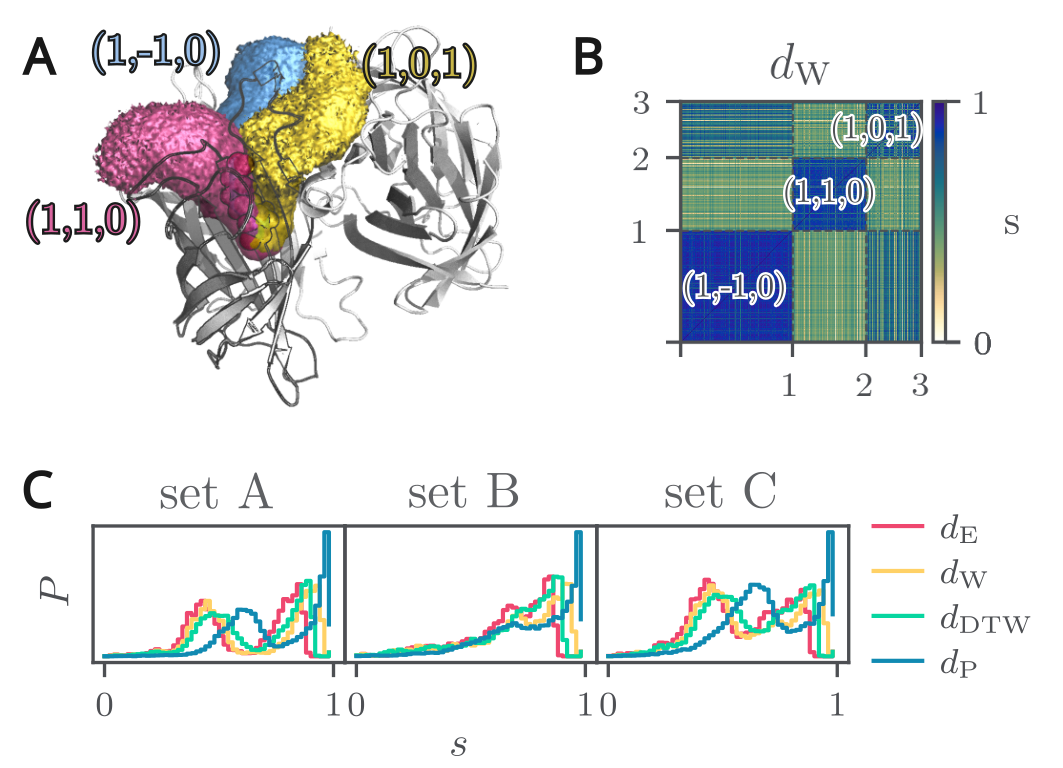}
    \caption{
    A: Rendering of streptavidin-biotin tetramer. Different biotin unbinding pathways are shown as translucent volumes. Streptavidin is represented as a cartoon.
    B: Block-ordered similarity matrix of constraint ground truth clusters.
    C: Distributions of pairwise similarities \( s_{i,j} \) for the constraint-pulling streptavidin-biotin test sets A, B, and C across different distance measures.
    }
    \label{fig:Stb_Panel1}
\end{figure}

In the following, we begin by testing different (Euclidean distance, DTW, Wasserstein and Procrustes-based) similarity measures for pathway clustering in the Streptavidin–biotin (St-b) system. This relatively straightforward test system with well-defined unbinding paths (see Fig.~\ref{fig:Stb_Panel1}) allows us to tune our clustering approach and to determine which data preprocessing strategies are most beneficial and to identify suitable parameter values for $\gamma$. After having established optimized parameters with St-b, we apply the same methods to the more demanding A2A-ZMA complex.

\subsection{Streptavidin-biotin test systems}

\begin{table*}[tb]
    \centering
    \caption{Comparison of computation times for different similarity measures on 12 threads of an AMD Ryzen 9 7950X 16-core CPU for 235 trajectories with 2001 time steps and 168 contact distances.}
    \label{tab:computation_times}
    \begin{tabular}{l|c|c|c}
    \textbf{distance metric} & \textbf{preprocessing} & \textbf{wall clock time} & \textbf{time order}\\
       & & & \textbf{dependence}\\
    \hline\hline
    Euclidean $d_{\rm E}$  & none              & 24~s & dependent\\
      & $\sigma$=5, t-norm, PC$_{1;4}$ & 16~s\\
    \hline
    DTW  $d_{\rm DTW}$      & none & $\approx\SI{26}{min}$&  dependent \\
            & $\sigma$=5, t-norm, PC$_{1;4}$ & 72~s &\\
    \hline
    Procrustes $d_{\rm P}$& none & $\approx\SI{11}{h}$ & dependent\\
     & $\sigma$=5, t-norm, PC$_{1;4}$ & $\approx\SI{11}{h}$                      \\
     \hline
    Wasserstein $d_{\rm W}$& none & 121~s & independent\\
     & $\sigma$=5, t-norm, PC$_{1;4}$ & 8~s                      \\
    \end{tabular}
\end{table*}

\paragraph{Computational cost of distance measures}
First, contact distance time traces were computed for all streptavidin-biotin (St-b) test sets.
Then, the following preprocessing steps were systematically applied to the contact distances: (i) temporal smoothing using a Gaussian filter with standard deviations $\sigma = 2, 5$, and 10 frames; (ii) normalization using either time-resolved scaling (Eq. \eqref{eq:normT}, labeled as n$_t$) or global normalization (Eq.~\eqref{eq:global_norm}, labeled as n) and (iii) dimensionality reduction via Principal Component Analysis. In previous work, we found that PCs 1-4 are a suitable subspace that contains most of the relevant information.\cite{Taenzel24}
Table \ref{tab:computation_times} displays the computation times for selected similarity preprocessing combinations.
Among the tested similarity measures, the fastest to compute is the Wasserstein distance $d_{\rm W}$ on PCA-reduced contact distances (PC1-4), with a wall clock time of $\Delta t \approx \SI{8}{s}$, followed by Euclidean distances $d_{\rm E}$ using the same preprocessing with $\Delta t \approx \SI{16}{s}$. 
Procrustes distances $d_{\rm P}$ are by far the most computationally expensive, as its calculation scales with the number of timesteps $K$ on the order of $\mathcal{O}(K^3)$:
computing the similarity matrix without any preprocessing took approximately 11 hours. To make a comparison with the other methods feasible, we downsampled the trajectories by including only every 5\textsuperscript{th} time step for the preprocessed contact distances, which reduced the wall clock time to $\Delta t \approx\SI{100}{s}$. The downsampling does not significantly influence the resulting similarities  (see Fig.~\SIprocrustessimilaritycomparison). In the following, all analyses concerning Procrustes are therefore performed with the downsampled data set. 
Dynamic Time Warping $d_{\rm DTW}$ when applied to PCA-reduced contact distances results in a similar calculation time of $\Delta t \approx\SI{72}{s}$.

We note that the PCA-induced dimensionality reduction of minimal contact distances is generally beneficial for the computational performance: employing the fully dimensional data increases the calculation time for Wasserstein distances and DTW by an order of magnitude. The Euclidean distances are relatively robust, with only a 1.5-fold increase. Interestingly, the computational cost of Procrustes analysis appears to be independent of the data dimensionality. 

\paragraph{Constraint simulations}
We now turn to evaluate the performance of the different distance measures for trajectory similarity calculations of the St-b benchmarking system, beginning with constant velocity constraint data. The resulting similarity distributions for all datasets are shown in Figs.~\ref{fig:Stb_Panel1}C and \SIconstraintsimilaritydistributioncomp. Set A exhibits a distinct bimodal distribution, indicating two clearly separated groups of trajectories. In set B, we generally observe higher similarity values, reflecting that the two subsets of trajectories follow more similar unbinding pathways. Set C exhibits a comparatively larger population of similarity values below 0.6, indicating a broader overlap of enforced paths and less pronounced separation between bundles of trajectories.
When comparing the different similarity measures, $d_{\rm E}$, $d_{\rm W}$, and $d_{\rm DTW}$ produce overall similar distribution shapes across all sets. The Procrustes-based similarity $d_{\rm P}$ exhibits a high density of values close to one and a pronounced secondary peak in sets A and C.

The impact of preprocessing on the similarity distributions is illustrated in Fig.~\SIconstraintsimilaritydistributioncomp. 
Both PCA and global normalization preserve the characteristic shape of the similarity distributions for Euclidean $d_{\rm E}$, Wasserstein $d_{\rm W}$ and DTW distances $d_{\rm DTW}$. In contrast, time-resolved normalization (n$_t$) alters the distribution more substantially. For Procrustes-based similarity $d_{\rm P}$, any form of preprocessing except for temporal smoothing leads to significant changes in the similarity distribution. This is expected, as Procrustes analysis is highly sensitive to matrix geometry, and transformations such as normalization or PCA distort the relative spatial arrangement that the method is designed to evaluate.

\begin{figure}[htb!]
    \centering
    \includegraphics[width=0.95\linewidth]{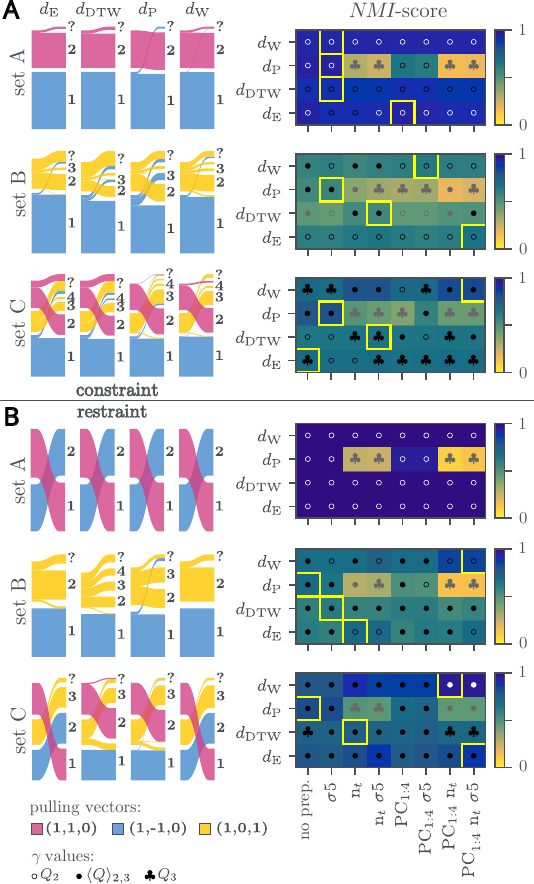}
    \caption{
    Clustering results of the streptavidin-biotin constraint (A) and restraint (B) pulling simulations with defined pulling directions. Left: Sankey diagrams for the highest normalized mutual information (NMI) score results for each measure for sets A (top), B (middle), and C (bottom). Small clusters with five or fewer members are grouped under the "?" label. Right: NMI scores for different similarity measures with different data preprocessing. Gaussian filtering is denoted by \( \sigma \), time normalization by \( n_t \), and PCA by PC\(_{1;4}\), indicating that principal components 1 through 4 were used. The symbol in the squares signifies the $\gamma$ values
    $Q_2$ ($\circ$), $\langle Q\rangle_{2,3}$ ($\bullet$) and $Q_3$ ($\clubsuit$) used for Leiden clustering resulting in a maximal NMI score. The color of the symbols was adjusted for improved visibility dependent on the underlying field color. The highest NMI scores for each similarity measure are marked with a yellow box. 
    }
    \label{fig:Stb_panel2}
\end{figure}

We clustered all computed similarity matrices using the Leiden algorithm and tested three values for the resolution parameter $\gamma$: the median ($Q_2$) of the respective similarity distribution, the third quartile ($Q_3$), and the average of the two ($\langle Q \rangle_{2,3}$).
Since the true pulling directions are known, we assess the clustering quality using normalized mutual information (NMI) scores. The results are summarized in Fig.~\ref{fig:Stb_panel2}A, comparing NMI scores for different similarity measures with varying data preprocessing and displaying Sankey diagrams for the highest NMI score results for each measure. Small clusters with five or fewer members are grouped together into a ''?'' cluster.  
NMI scores for all tested preprocessing configurations and $\gamma$ values are provided in Figs.~\SIStbconstraintsetAmiscoremat\ to \SIStbconstraintsetCmiscoremat.

In set A, which is the most easily separable, all similarity measures achieve NMI scores of close to one when clustered with a resolution parameter $\gamma=Q_2$. The slight deviation from one is due to the emergence of small clusters containing fewer than five trajectories. The only exception in this set are Procrustes-based distances, where applying preprocessing steps other than smoothing significantly degrade the NMI score.

For set B, the best overall clustering was obtained using Euclidean distance with n$_t$ and smoothing ($\sigma = 5$), followed by Wasserstein distance (also with n$_t$) and Procrustes similarity. These results were obtained using $\gamma = Q_2$ or $\langle Q \rangle_{2,3}$.
DTW achieved its highest scores (NMI = 0.61)
when preprocessing with n$_t$ and smoothing ($\sigma = 5$) was applied, along with a resolution parameter $\gamma = \langle Q \rangle_{2,3}$. However, lower NMI scores in this test set were due to trajectories in cluster (1,0,1) distributed across multiple small clusters. Of the four tested similarity measures, $d_{\rm W}$ recovers the largest cluster composed exclusively of trajectories from (1,0,1). DTW performs the worst: a substantial portion of (1,0,1) trajectories is split into many small clusters and ends up in "?".

Set C consists of trajectories from three distinct pulling directions. These yield geometrically well-defined but partially overlapping clusters in feature space. Clustering this set poses greater challenges since the trajectories corresponding to (1,0,1) and (1,-1,0) have a higher inter cluster similarity than (1,1,0) and (1,-1,0). 
The highest NMI scores are observed for $d_{\rm W}$ applied to preprocessed contact distances (smoothing, n$_t$ and PCA) and Procrustes based similarities using $\gamma = \langle Q \rangle_{2,3}$.

\paragraph{Restraint simulation} clustering results are shown in 
Fig.~\ref{fig:Stb_panel2}B, and similarity distributions in Fig.~\SIrestraintsimilaritydistributioncomp.
Interestingly, the clustering outcomes generally yield higher NMI scores than the constraint simulations. This is likely because the ligand can follow a more natural unbinding pathway when restraints are used instead of constraints, which often impose hard geometric boundaries on the dissociation process. 

For set A, all tested similarity measures with various preprocessing methods achieved perfect normalized mutual information scores (NMI = 1) across all distance metrics, again with the exception of preprocessed Procrustes analysis-based similarities. 

In set B, the best clustering result is achieved using $d_{\rm W}$ in combination with time normalization and PCA. This configuration yields an NMI score of 0.88 with $\gamma=Q_2$. Procrustes-based clustering ranks second but splits the trajectories in (1,0,1) into two separate clusters. The worst performance is observed for $d_{\rm DTW}$, which fragments (1,0,1) into multiple clusters and introduces minor mixing with (1,-1,0). $d_{\rm E}$ performs better than $d_{\rm DTW}$ as it  successfully identifies one dominant cluster composed of (1,0,1) trajectories, but it introduces minor mixing with (1,-1,0)

For set C, path separation proved more straightforward than in the corresponding constraint simulations. The highest score (NMI > 0.85) is observed for $d_{\rm W}$ again after preprocessing with time-normalization and PCA, while applying additional smoothing yielded a comparable NMI score. $d_{\rm E}$ also performed well using the same preprocessing and successfully recovered all three ground-truth clusters. Only minor mixing was observed between (1,0,1) and (1,$-1$,0), and a small subset of trajectories from (1,0,1) is assigned to small clusters containing $n < 5$ trajectories. The remaining two similarity measures introduced in substantially more mixing between (1,0,1) and (1,$-1$,0).

Based on these results, we selected the preprocessing parameters that yielded the highest clustering scores on sets B and C for each distance type for downstream applications, specifically:
\begin{itemize}
    \item DTW with n$_t$ and $\sigma=10$,
    \item Wasserstein distances with PCA, n$_t$ and $\sigma=5$,
    \item Procrustes without preprocessing,
    \item Euclidean distances without preprocessing.
\end{itemize}
In the further analyses (e.g., A\down{2a} clustering), we apply these optimal configurations.

\subsection{A\textsubscript{2a} receptor}

\begin{figure}[htb!]
    \centering
    \includegraphics[width=0.95\linewidth]{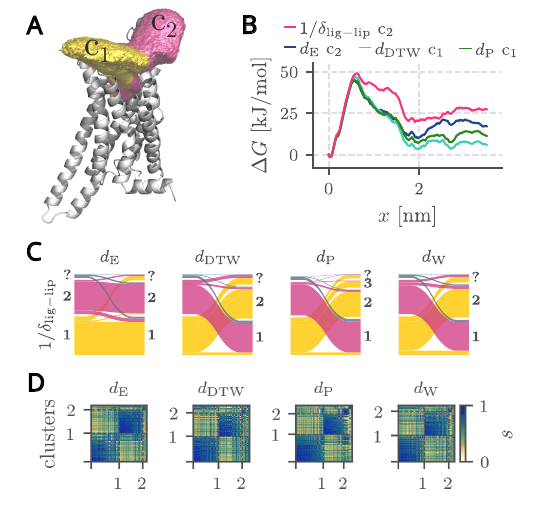}
    \caption{Clustering results for the A2a adenosine receptor-inhibitor complex using $\gamma=Q_2$ compared to known clusterig results based on the inverse minimal ligand-lipid distance $1/\delta_\text{lig-lip}$ as a reaction coordinate with microscopically feasible unbinding mechanism.
    A: Visualization of cluster 1 and 2 as volumes based on inverse ligand lipid distances
    B: pathwise free energies. C: sankey diagram comparing geometrical path separation results with $1/\delta_\text{lig-lip}$ D: block ordered similarity matrix.}
    \label{fig:A2A_panel1}
\end{figure}

Following the benchmarking on streptavidin-biotin, we apply our workflow to the more challenging case of a thermostabilized variant of the A\textsubscript{2a} receptor bound to the antagonist ZM241385,\cite{Segala2016} 
where we enforce ligand unbinding via a constant velocity constraint.
From a previous study\cite{Taenzel24} we known that unbinding pathways in this system can be characterized by the distance between the ligand and membrane lipids. The two major paths are visualized in Fig.~\ref{fig:A2A_panel1}A. In the following these microscopic property informed pathways will be referred to as ground truth. In addition to the comparison of clusterings, we derive free energy profiles $\Delta G(x)$ for each cluster containing more than 100 trajectories via dcTMD and compare them to ground truth reference free energies in Fig.~\ref{fig:A2A_panel1}B.
We use Leiden/CPM clustering with $\gamma$ values equal to the median of the pairwise similarities ($Q_2 = \text{median}(s_{ij})$, see Fig.~\ref{fig:A2A_panel1}C,D)
and the third quartile ($Q_3$, see Fig.~\SIAAPanelQ). Clustering the similarity matrices with $\gamma=Q_2$ results in two dominant clusters for all measures (Fig.~\ref{fig:A2A_panel1}D), which agrees well with our ground truth.

Fig.~\ref{fig:A2A_panel1}C visualizes the correspondence between the ground-truth and similarity-based clusters using Sankey diagrams. All four similarity measures display strong consistency with the ground truth, with the Euclidean distance $d_{\rm E}$ showing the least mixing of trajectories belonging to ground truth cluster $c_2$.

At this point in our investigation, we have to remark that identifying pathways in biased MD simulations does not yield information about their meaningfulness under equilibrium conditions, i.e., if transitions between bound and unbound states indeed follow these paths, or if they are only artifacts of the biasing protocol. However, it is possible to extract such information, e.g., from estimates of kinetics in infrequent Metadynamics\cite{Tiwary15}, by dcTMD and pathway probability reweighing\cite{Wolf2023} or via using path-representative trajectories for transition pathway sampling and committor analysis.\cite{Jung23,Megias2025}
In the framework of our work, we applied dcTMD to calculate the pathwise free energy for each cluster.
In agreement with our ground truth clusters, only one of the two clusters of each respective distance measure yields a physically plausible free energy profile in which the bound state is energetically more favorable than the unbound state, except for $d_{\rm W}$, where no physically meaningful cluster is found when $\gamma=Q_2$ is used.
Figs.~\ref{fig:A2A_panel1}B and \SIAAgeometricdG\ display these free energy profiles. 
For $d_{\rm W}$ no cluster physically meaningful cluster, i.e., an unbound state that is has a lower free energy than the unbound state, is found when $\gamma=Q_2$ is used, which is a hallmark of an unsuccessful pathway separation.\cite{Wolf2023} All three remaining measures yield free energy barriers of $\Delta G^\ddagger \approx 45$--$50\,\mathrm{kJ/mol}$. The unbound-state free energy varies by metric: the best agreement with the reference $\approx 25\,\mathrm{kJ/mol}$ is given for $d_{\rm E}$ with $\approx 20\,\mathrm{kJ/mol}$, followed by $d_{\rm P}$ with $\approx 10\,\mathrm{kJ/mol}$. The last position is held by $d_{\rm DTW}$ with only $\approx 5\,\mathrm{kJ/mol}$.

Mitigation of this deviation between ground truth and predicted free energy profiles is achieved by increasing the resolution factor $\gamma$. Further examination of cluster composition with $\gamma = Q_3$ results in splitting and remixing of the two dominant clusters into smaller, purer sub-clusters (see Fig.~\SIAAPanelQ) In contrast to $\gamma = Q_2$, the two main clusters are mostly kept homogeneously, albeit at the cost of a $\sim$40\% reduction in cluster size. The trajectories removed from this set mix together into the novel third and additional smaller clusters.
Concerning the resulting free energy estimates as displayed in Fig.~\SIAAPanelQ, $d_{\rm W}$ now yields a reasonable free energy estimate. The barrier height increases only moderately by $5$--$10\,\mathrm{kJ/mol}$ for all investigated similarity measures, while the unbound free energy rises to $20-30\,\mathrm{kJ/mol}$. The best agreement with the ground truth is given for $d_{\rm E}$, $d_{\rm W}$ and $d_{\rm DTW}$, while $d_{\rm P}$ still underestimate the unbound state free energy by $\approx 10\,\mathrm{kJ/mol}$. 

Given A\down{2a} as realistic benchmark system, we thus recommend to use $d_{\rm E}$- and $d_{\rm W}$-based similarities in combination with the Leiden-CPM community analysis in a realistic investigation of unbinding paths. If bias information along the paths can be used to infer free energies, e.g., by dcTMD as done here, we further recommend to start with $\gamma = Q_2$ and stepwise increase to $\gamma = Q_3$ or even beyond, checking free energy convergence, while ensuring that the investigated clusters contain at least 100 trajectories each.\cite{Wolf2020,Taenzel24}
As the normality plots in Figs.~\SIAAfigoneworkQQ\ and \SIAAfigtwoworkQQ\ display, the work distribution along all resulting physically meaningful paths is well comparable to a normal distribution.
We note that because of the persisting friction overestimation for the pathway along the membrane-solvent interface, we currently cannot state anything about the physical feasibility of both pathways. To provide this information, we will carry out temperature-boosted Langevin simulations with different A\down{2a} ligands and compare them with experimental rates in future works.

\section{Conclusions}

In this work, we systematically benchmarked four trajectory similarity measures - Euclidean, Wasserstein, Procrustes, and Dynamic Time Warping (DTW) - for their capacity to classify unbinding pathways of protein-ligand complexes from biased molecular dynamics simulations. Using the streptavidin-biotin system as a ground-truth benchmark and the A\textsubscript{2a} receptor-antagonist complex as a realistic application case, we demonstrated that
more sophisticated measures than simple Euclidean distances apparently are not necessarily better at sorting trajectory ensembles according to pathways. Especially DTW and Procrustes exhibit exceedingly expensive computational cost, but do not provide more information over $d_{\rm E}$ or even come at the price of degraded performance in the case of $d_{\rm P}$. We note that DTW may indeed perform better for methods with strongly different time traces such as Metadynamics,\cite{Ray24} we do not see any hint for this in both the geometrically strict constraint simulations and the more flexible steered MD simulations. For a reasonable free energy estimate via dcTMD, we recommend to test setting $\gamma = Q_3$ instead of $Q_2$ as recommended by us earlier, which however may come at the expense of smaller trajectory clusters and thus smaller sampling.

The Wasserstein distance $d_{\rm W}$ performance improved substantially when combined with principal component analysis (PCA) and time normalization (n\textsubscript{t}), resulting in higher NMI scores than $d_{\rm E}$. It is surprisingly good for a method that removes time dependence completely during evaluation. Apparently the distribution of contact probabilities is more important for path discrimination than the detailed time information. This is in qualitative agreement with interaction fingerprints performing well for path separation.\cite{Kokh20}

Closing this work with a practical guideline for pathway detection via trajectory clustering, we recommend to use either Euclidean distances $d_{\rm E}$ on the full protein-ligand contact set, or $d_{\rm W}$ preferably on some pre-processed data of suitable input features for inter-trajectory similarity calculations. Both methods represent the optimum between computational cost and accuracy of the resulting similarities. The Leiden/CPM clustering proves to be a practical and effective method for the clustering of biased simulation trajectories. Its performance is robust to moderate variation in the resolution parameter $\gamma$, and it requires minimal user input, especially no previous knowledge about the number of trajectory clusters and thus pathways existing. We note that our results do not only have implications for the pathway sorting of data from biased simulations, but for the preparation of common equilibrium-based or stationary free energy calculations, as it e.g. has been recently shown that Umbrella sampling calculations on the dissociation paths of biomolecular complexes yield wrong results when not taking into account multiple possible unbinding paths.\cite{Aho24}
While the relative performance of distance metrics should be fairly robust across biasing protocols, the biophysical plausibility of the recovered pathways may of course vary between methods. On the other hand, the main conclusions of the work could be valid also for spontaneous (unbiased) unbinding if it occurs on a computationally accessible timescale.\cite{Greisman23}

\begin{acknowledgement}

The authors thank the Deutsche Forschungsgemeinschaft (DFG, German Research Foundation) for the grant WO 2410/2-1 within the framework of the Research Unit FOR 5099 ''Reducing complexity of non-equilibrium systems'' (Project No. 431 945 604) for financial support. They acknowledge support by the bwUniCluster computing initiative, the High Performance and Cloud Computing Group at the Zentrum für Datenverarbeitung of the University of Tübingen, the state of Baden-Württemberg through bwHPC, and the DFG through Grant No. INST 37/935-1 FUGG. They are further grateful to Gerhard Stock, Georg Diez, Daniel Nagel, Sofia Sartore and Victor T\"anzel for helpful discussions.
shown in this document.

\end{acknowledgement}

\begin{suppinfo}

Information and one Supplementary Table on the simulation details of the streptavidin-biotin simulation system as well as 10 Supplementary Figures on similarity distributions, NMI scores, alternative clusterings of A\down{2a} and free energy estimates by dcTMD. (PDF)

\end{suppinfo}

\section*{Author declarations}

\subsection*{Conflict of Interest}

The authors declare no conflict of interest.

\subsection*{Author contributions}

Both authors designed the research project. MJ carried out research, SW supervised the project. MJ prepared all Figures and Tables and wrote the original draft. Both authors reviewed and edited the manuscript.

\section*{Data Availability Statement}

The data that support the findings of this study are available from the corresponding author upon reasonable request.

\bibliography{bib_new_2025}

\newpage

\onecolumn

\renewcommand{\thepage}{\arabic{chapter}.\arabic{page}}  
\renewcommand{\thesection}{\arabic{chapter}.\arabic{section}}   
\renewcommand{\thetable}{\arabic{chapter}.\arabic{table}}   
\renewcommand{\thefigure}{\arabic{chapter}.\arabic{figure}}
\renewcommand{\thepage}{S\arabic{page}}  
\renewcommand{\thesection}{S\arabic{section}}   
\renewcommand{\thetable}{S\arabic{table}}   
\renewcommand{\thefigure}{S\arabic{figure}}

\setcounter{figure}{0}
\setcounter{table}{0}

\section*{Supporting Information}

\subsection*{Streptavidin-Biotin Simulation Details}

The simulations were carried out using different pulling vectors in space with a Cartesian $(x, y, z)$ nomenclature.
For each pulling direction, 50 restraint simulations and 200 constraint simulations were started, respectively. Table~\ref{tab:numer_traj} shows the pulling directions with the corresponding number of successful trajectories. Unsuccessful trajectories resulted from pulling the ligand unphysically into the protein and causing a simulation ''crash'' due to too high interparticle forces.

\begin{table}[h!]
\centering
\begin{tabular}{c|c|c}

\textbf{Pulling Vector} & \multicolumn{2}{c}{Number of trajectories}  \\
\hline 
 & constraint & restraint \\
\hline\hline
(1, 0, 1)   & 71 & 40 \\
(1, 1, 0)   & 93  & 50 \\
(1, -1, 0)  & 142 & 48 \\

\end{tabular}
\caption{Number of successful trajectories for each pulling vector in constraint and restraint simulations.}
\label{tab:numer_traj}
\end{table}

\subsection*{Performance of Similarity Measures}
\begin{figure}
    \centering
    \includegraphics[width=0.5\linewidth]{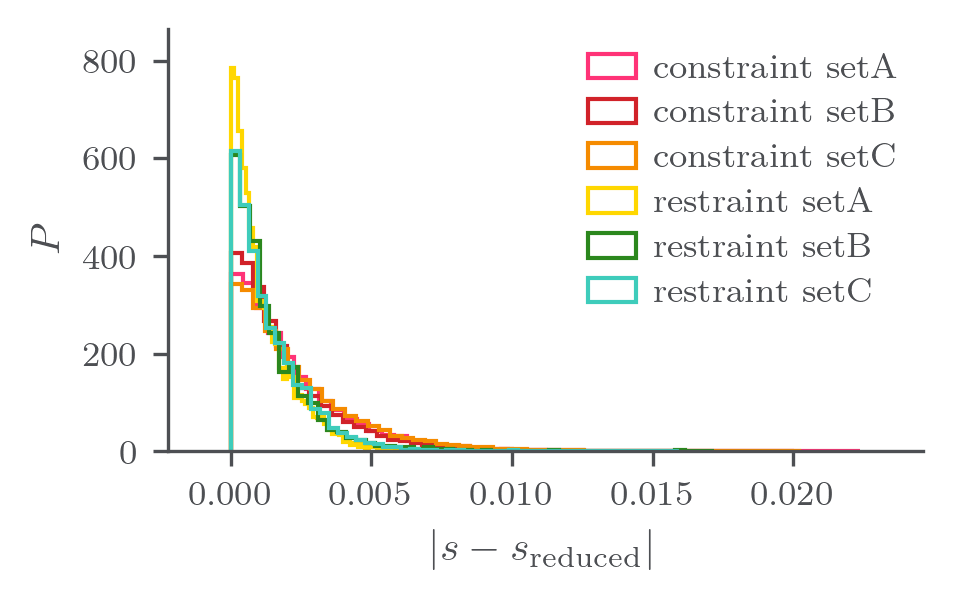}
    \caption{Distribution of the absolute difference between the similarities calculated from the full trajectories ($s$) and the downsampled trajectories ($s_\text{reduced}$) of the different streptavidine biotine test sets.}
    \label{fig:SIprocrustes_downsampling_similarity_difference}
\end{figure}
\begin{figure}[h!]
    \centering
    \includegraphics[width=0.8\linewidth]{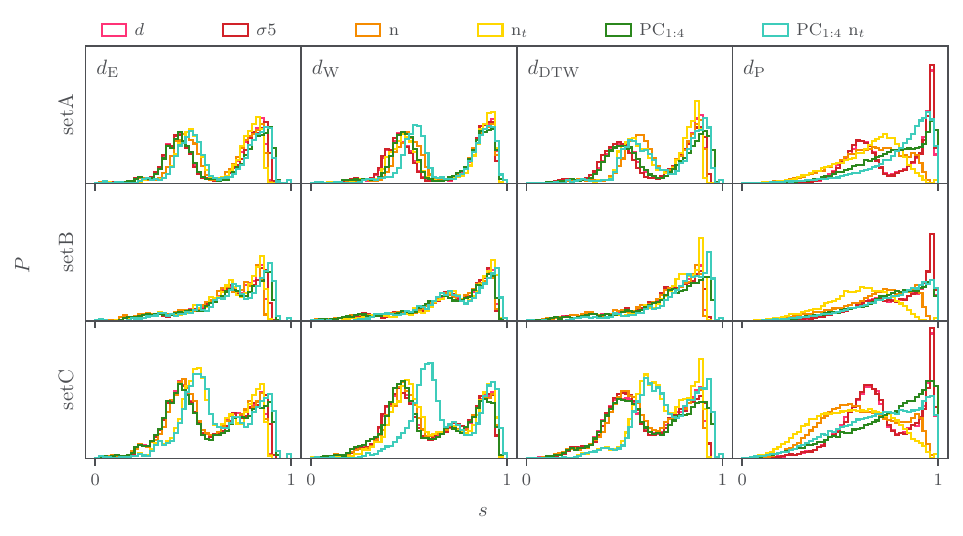}
    \caption{Comparison of the influence of preprocessing on the similarity distribution for streptavidin-biotin constraint simulations sets A, B and C. The similarity measures is given by the label inside the Figure. The data preprocessing is given as color code in the legend. }      \label{fig:SIconstraint_similarity_distribution_comp}
\end{figure}

\begin{figure}[h!]
    \centering
    \includegraphics[width=0.8\linewidth]{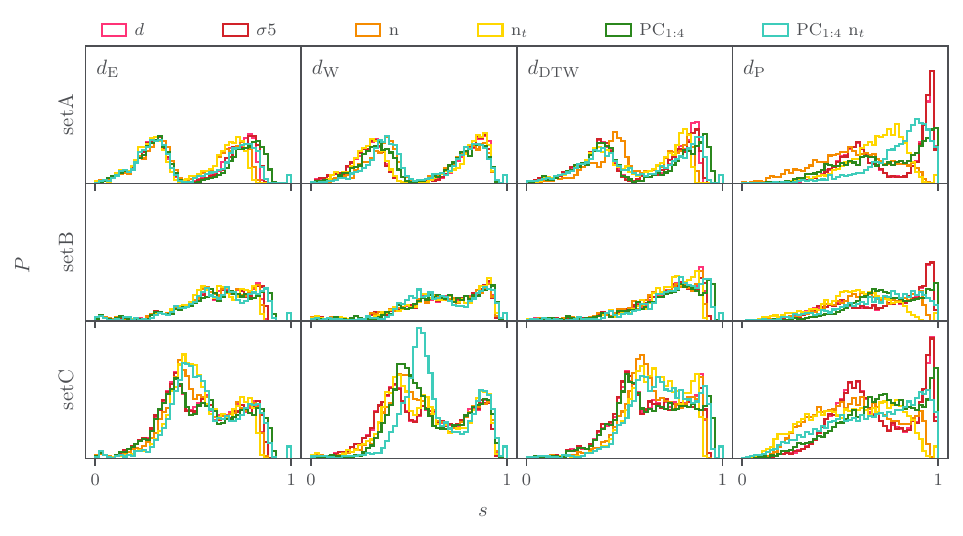}
    \caption{Comparison of the influence of preprocessing on the similarity distribution for streptavidin-biotin restraint simulations sets A, B and C. The similarity measures is given by the label inside the Figure. The data preprocessing is given as color code in the legend. }    \label{fig:SIrestraint_similarity_distribution_comp}
\end{figure}

\begin{figure}[h!]
    \centering
    \includegraphics[width=0.8\linewidth]{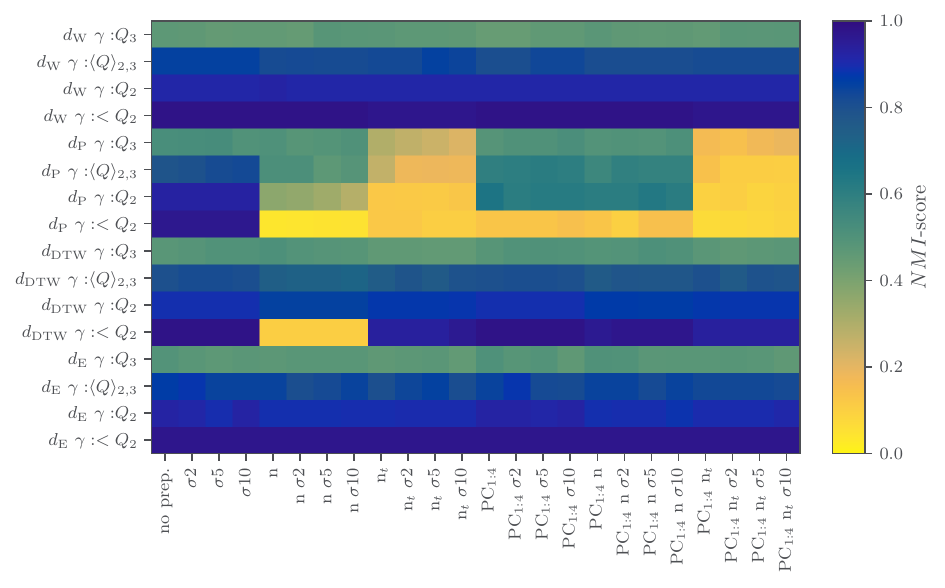}
    \caption{
        Comparison of clustering results from streptavidin-biotin constraint simulations set A with ground truth clusters using the NMI score. The $y$-axis indicates the similarity measure used and the resolution parameter $\gamma$: $Q_2$ represents the median of the similarity distribution, $Q_3$ the third quartile, $\langle Q \rangle_{2,3}$ the mean of $Q_2$ and $Q_3$, and $<Q_2$ denotes $Q_2 - 0.1$. The $x$-axis indicates the preprocessing applied to the dataset before similarity calculation: $\sigma$ specifies the Gaussian filter width, n denotes global normalization, n$_t$ indicates time-resolved normalization and PC$_{1:4}$ refers to PCs 1-4 were used.
    }
    \label{fig:SI_Stb_constraint_setA_miscore_mat}
\end{figure}

\begin{figure}[h!]
    \centering
    \includegraphics[width=0.8\linewidth]{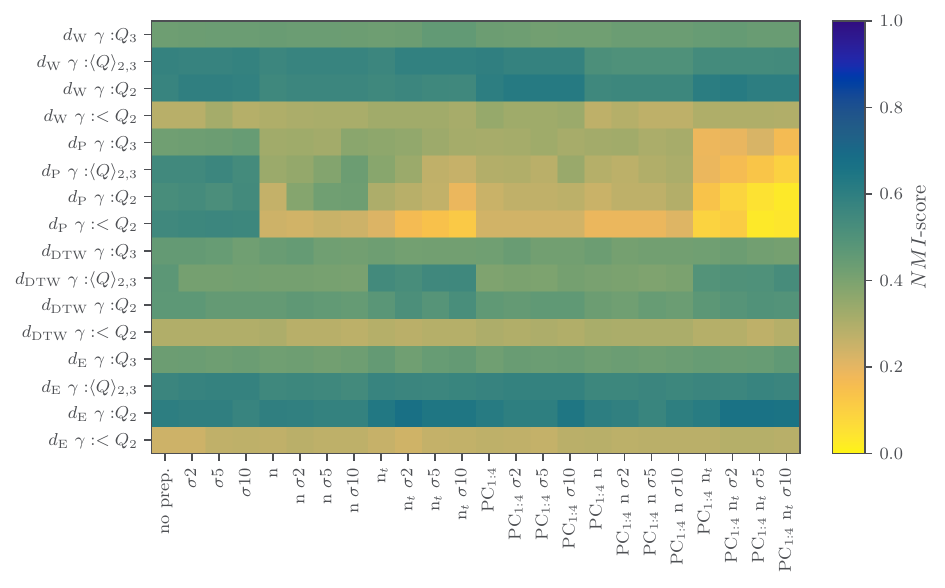}
    \caption{Comparison of clustering results from streptavidin-biotin constraint simulations set B with ground truth clusters using the NMI score. The $y$-axis indicates the similarity measure used and the resolution parameter $\gamma$: $Q_2$ represents the median of the similarity distribution, $Q_3$ the third quartile, $\langle Q \rangle_{2,3}$ the mean of $Q_2$ and $Q_3$ and $<Q_2$ denotes $Q_2 - 0.1$. The $x$-axis indicates the preprocessing applied to the dataset before similarity calculation: $\sigma$ specifies the Gaussian filter width, n denotes global normalization, n$_t$ indicates time-resolved normalization and PC$_{1:4}$ refers to PCs 1-4 were used.}
    \label{fig:SI_Stb_constraint_setB_miscore_mat}
\end{figure}

\begin{figure}[h!]
    \centering
    \includegraphics[width=0.8\linewidth]{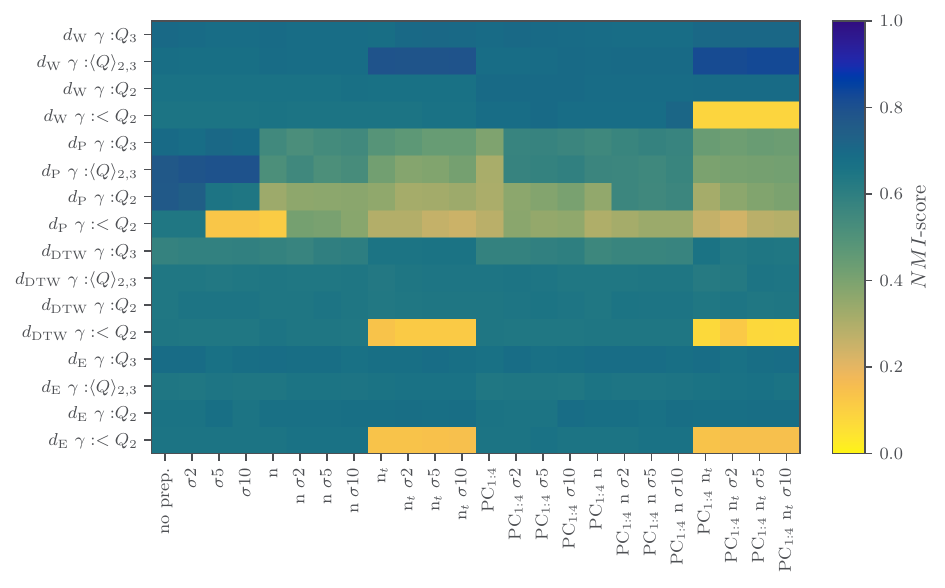}
    \caption{Comparison of clustering results from streptavidin-biotin constraint simulations set C with ground truth clusters using the NMI score. The $y$-axis indicates the similarity measure used and the resolution parameter $\gamma$: $Q_2$ represents the median of the similarity distribution, $Q_3$ the third quartile, $\langle Q \rangle_{2,3}$ the mean of $Q_2$ and $Q_3$, and $<Q_2$ denotes $Q_2 - 0.1$. The $x$-axis indicates the preprocessing applied to the dataset before similarity calculation: $\sigma$ specifies the Gaussian filter width, n denotes global normalization, n$_t$ indicates time-resolved normalization and PC$_{1:4}$ refers to PCs 1-4 were used.}
    \label{fig:SI_Stb_constraint_setC_miscore_mat}
\end{figure}

\begin{figure}[h!]
    \centering
    \includegraphics[width=0.8\linewidth]{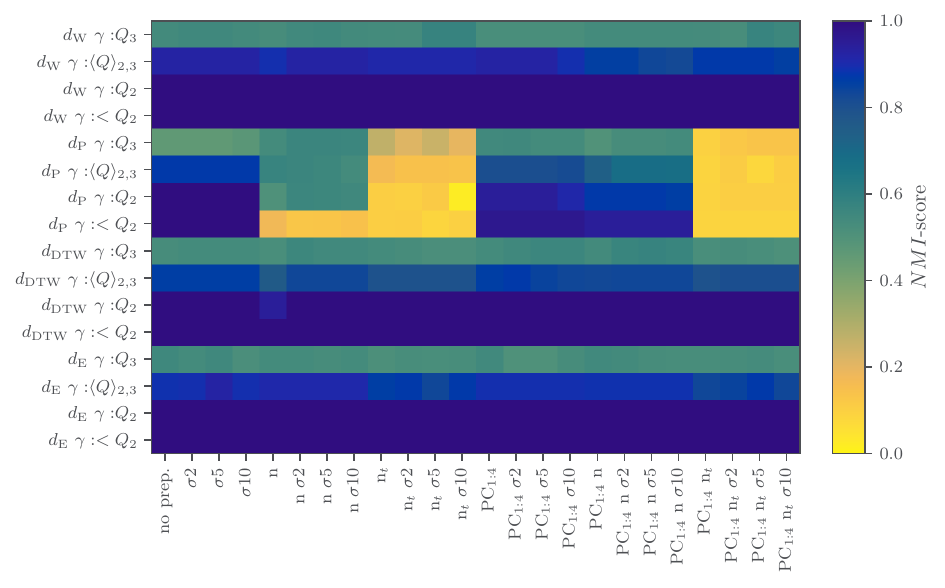}
    \caption{Comparison of clustering results from streptavidin-biotin restraint simulations set A with ground truth clusters using the NMI score. The $y$-axis indicates the similarity measure used and the resolution parameter $\gamma$: $Q_2$ represents the median of the similarity distribution, $Q_3$ the third quartile, $\langle Q \rangle_{2,3}$ the mean of $Q_2$ and $Q_3$, and $<Q_2$ denotes $Q_2 - 0.1$. The $x$-axis indicates the preprocessing applied to the dataset before similarity calculation: $\sigma$ specifies the Gaussian filter width, n denotes global normalization, n$_t$ indicates time-resolved normalization and PC$_{1:4}$ refers to PCs 1-4 were used.}
    \label{fig:SI_Stb_restraint_setA_miscore_mat}
\end{figure}

\begin{figure}[h!]
    \centering
    \includegraphics[width=0.8\linewidth]{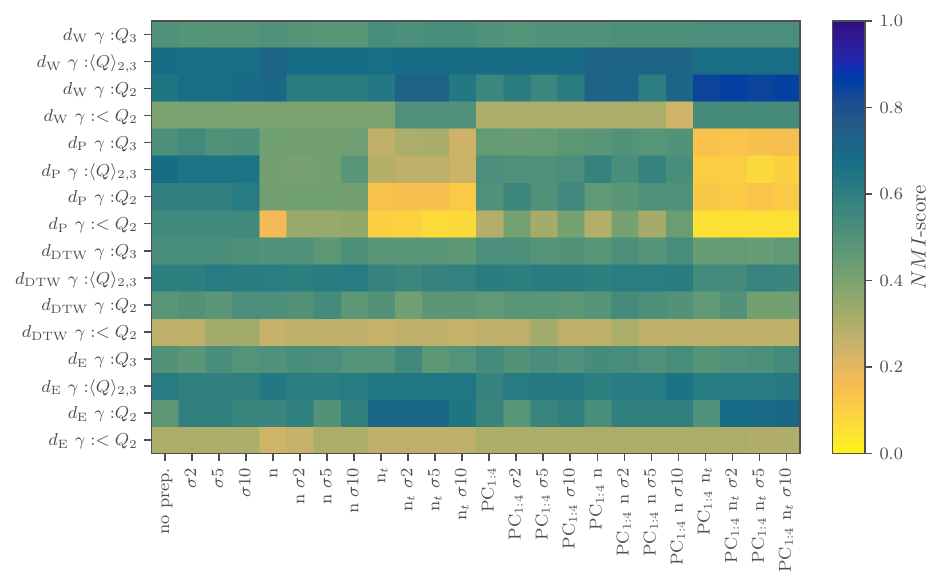}
    \caption{Comparison of clustering results from streptavidin-biotin restraint simulations set B with ground truth clusters using the NMI score. The $y$-axis indicates the similarity measure used and the resolution parameter $\gamma$: $Q_2$ represents the median of the similarity distribution, $Q_3$ the third quartile, $\langle Q \rangle_{2,3}$ the mean of $Q_2$ and $Q_3$, and $<Q_2$ denotes $Q_2 - 0.1$. The $x$-axis indicates the preprocessing applied to the dataset before similarity calculation: $\sigma$ specifies the Gaussian filter width, n denotes global normalization, n$_t$ indicates time-resolved normalization and PC$_{1:4}$ refers to PCs 1-4 were used.}
    \label{fig:SI_Stb_restraint_setB_miscore_mat}
\end{figure}

\begin{figure}[h!]
    \centering
    \includegraphics[width=0.8\linewidth]{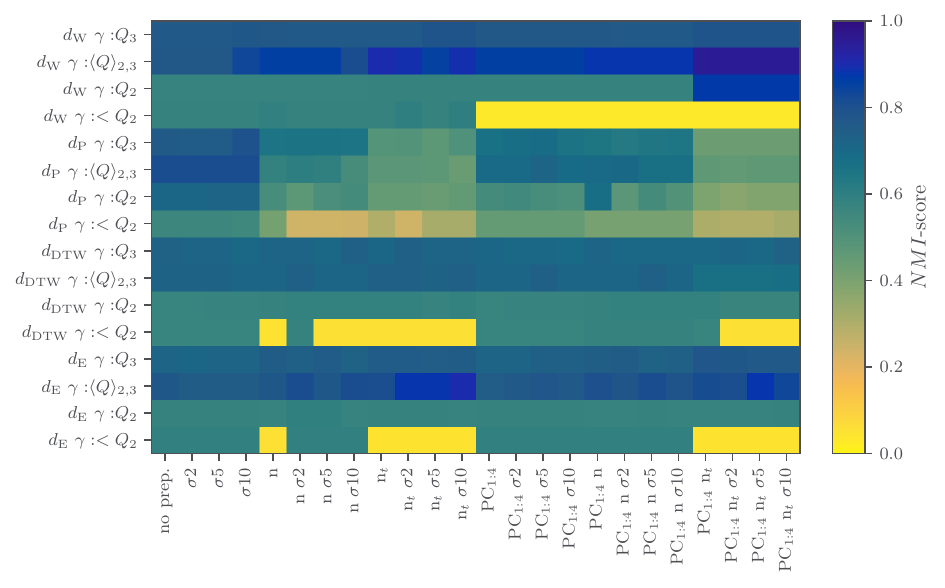}
    \caption{Comparison of clustering results from streptavidin-biotin restraint simulations set C with ground truth clusters using the NMI score. The $y$-axis indicates the similarity measure used and the resolution parameter $\gamma$: $Q_2$ represents the median of the similarity distribution, $Q_3$ the third quartile, $\langle Q \rangle_{2,3}$ the mean of $Q_2$ and $Q_3$, and $<Q_2$ denotes $Q_2 - 0.1$. The $x$-axis indicates the preprocessing applied to the dataset before similarity calculation: $\sigma$ specifies the Gaussian filter width, n denotes global normalization, n$_t$ indicates time-resolved normalization and PC$_{1:4}$ refers to PCs 1-4 were used.}
    \label{fig:SI_Stb_restraint_setC_miscore_mat}
\end{figure}

\begin{figure}[h!]
    \centering
    \includegraphics[width=0.8\linewidth]{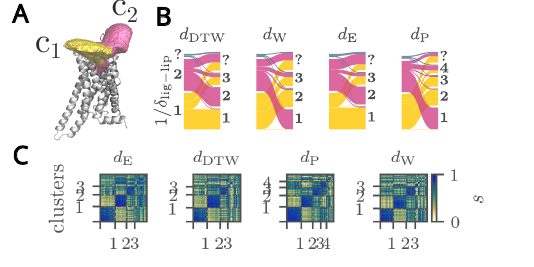}
    \caption{Clustering results for the A\down{2a} adenosine receptor-inhibitor complex using $\gamma=Q_3$ compared to known clustering results based on the inverse minimal ligand-lipid distance $1/\delta_\text{lig-lip}$ as a reaction coordinate with microscopically feasible unbinding mechanism.
    A: Visualization of cluster 1 and 2 as volumes based on inverse ligand lipid distances
    B: Sankey diagram comparing geometrical path separation results with $1/\delta_\text{lig-lip}$ clustered with $\gamma=Q_2$. C: Block ordered similarity matrix.}
    \label{fig:SI_A2A_PanelQ3}
\end{figure}

\begin{figure}[h!]
    \centering
    \includegraphics[width=0.8\linewidth]{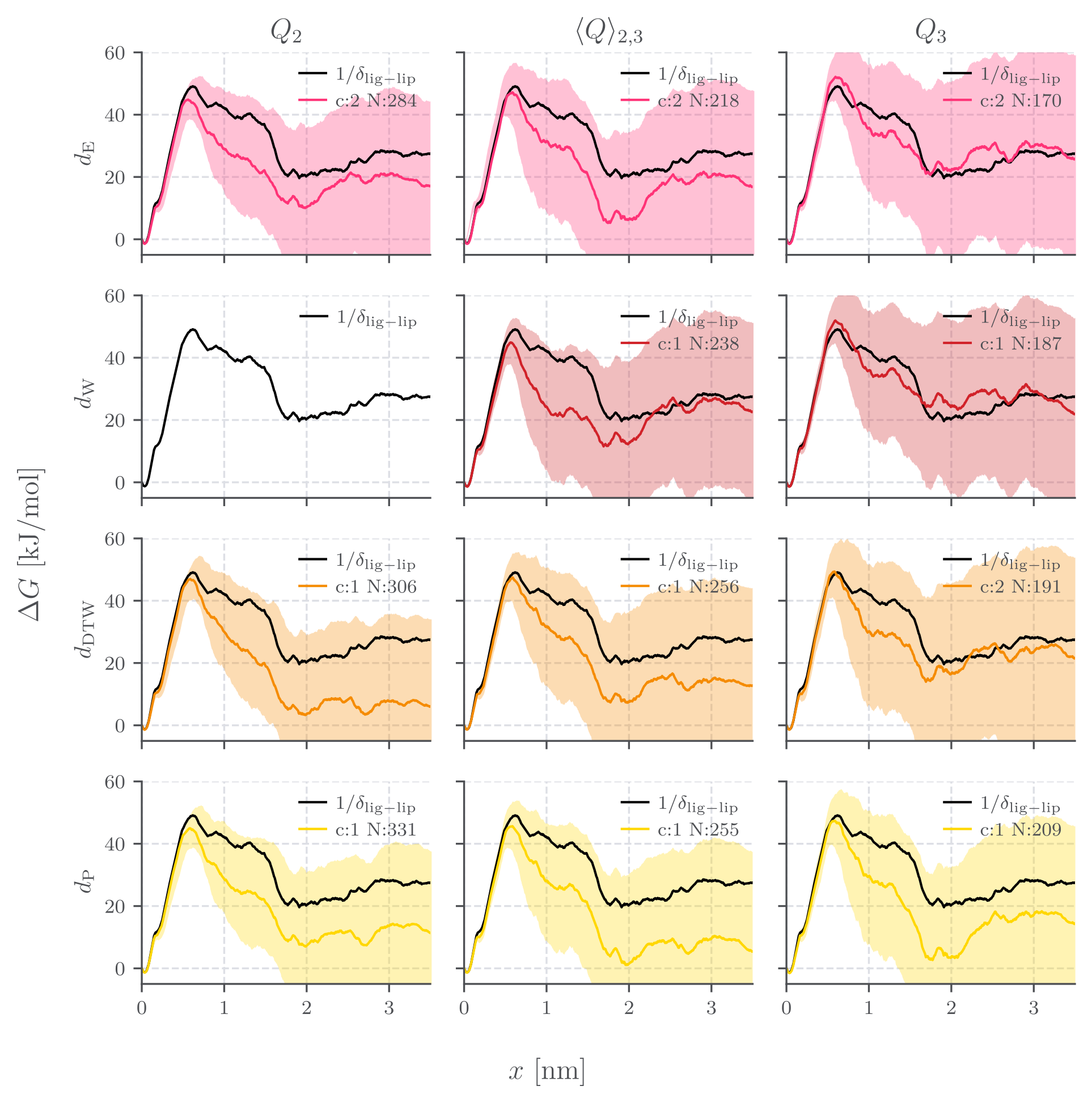}
    \caption{dcTMD results or different A\down{2a} clusters. The shaded area signifies the bootstrapping errors from 5000 resamples in a 90\% confidence interval.}
    \label{fig:SI_A2A_geometric_dG}
\end{figure}

\begin{figure}[h!]
    \centering
    \begin{subfigure}{0.49\linewidth}
        \includegraphics[width=0.9\linewidth]{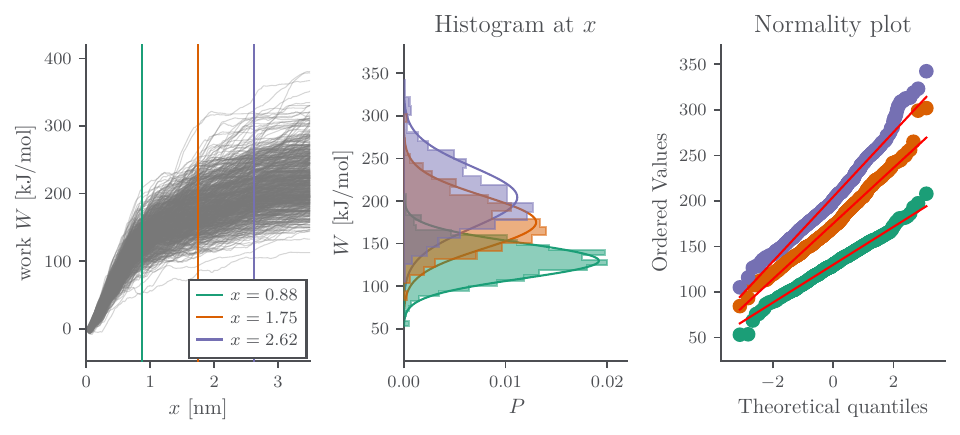}
        \caption{all trajectories}
    \end{subfigure}
    \begin{subfigure}{0.49\linewidth}
        \includegraphics[width=0.9\linewidth]{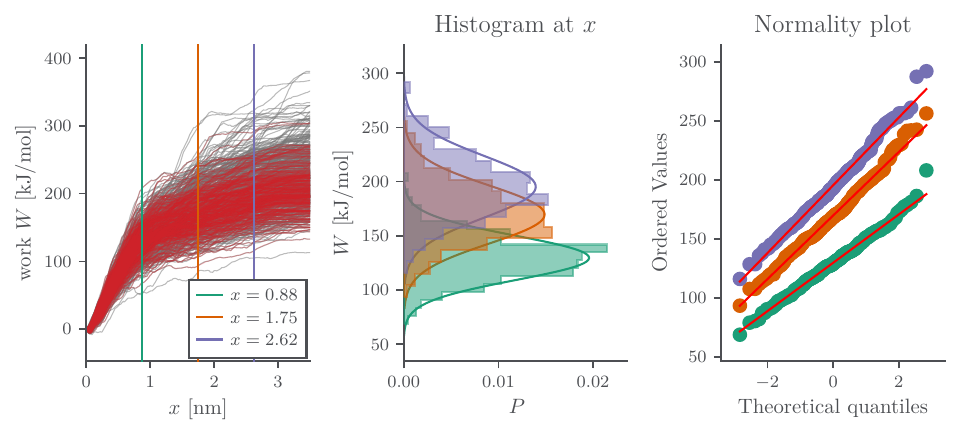}
        \caption{$1/\delta_\text{lig-lip}$ $c_2$ $\gamma=Q_2$}
    \end{subfigure}
    \begin{subfigure}{0.49\linewidth}
        \includegraphics[width=0.9\linewidth]{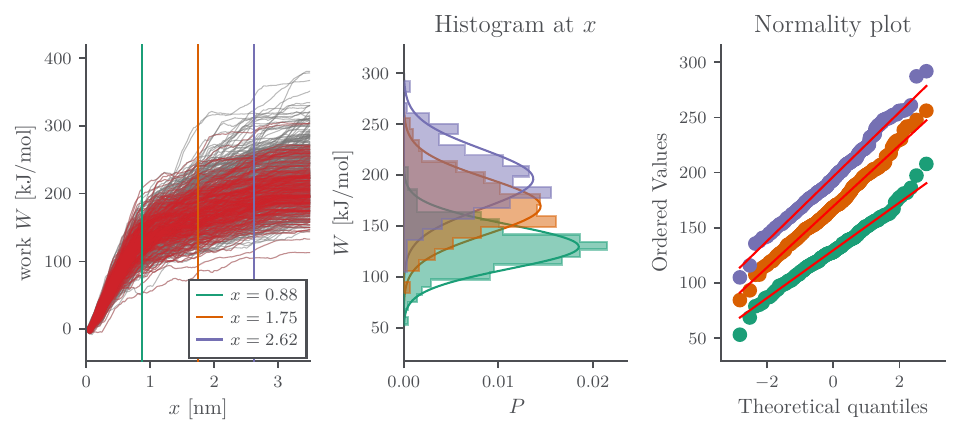}
        \caption{$d_\text{E}$ $c_2$ $\gamma=Q_2$}
    \end{subfigure}
    \begin{subfigure}{0.49\linewidth}
        \includegraphics[width=0.9\linewidth]{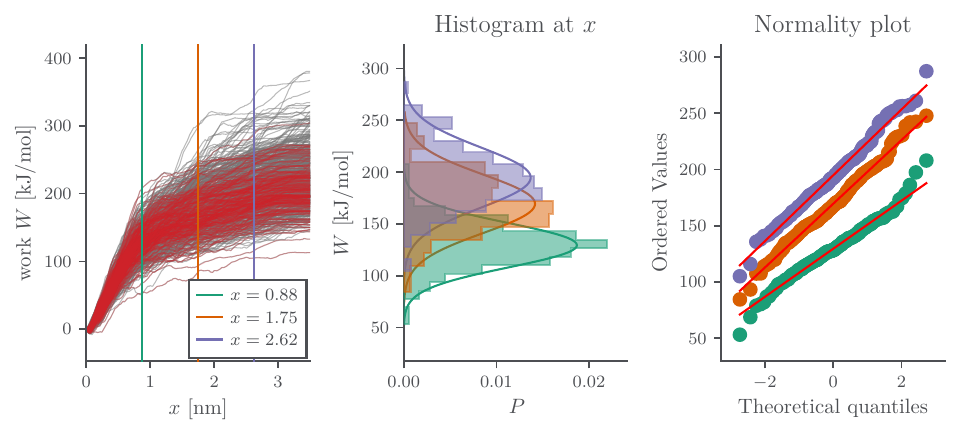}
        \caption{$d_\text{E}$ $c_2$ $\gamma=\langle Q\rangle_{23}$}
    \end{subfigure}
    \begin{subfigure}{0.49\linewidth}
        \includegraphics[width=0.9\linewidth]{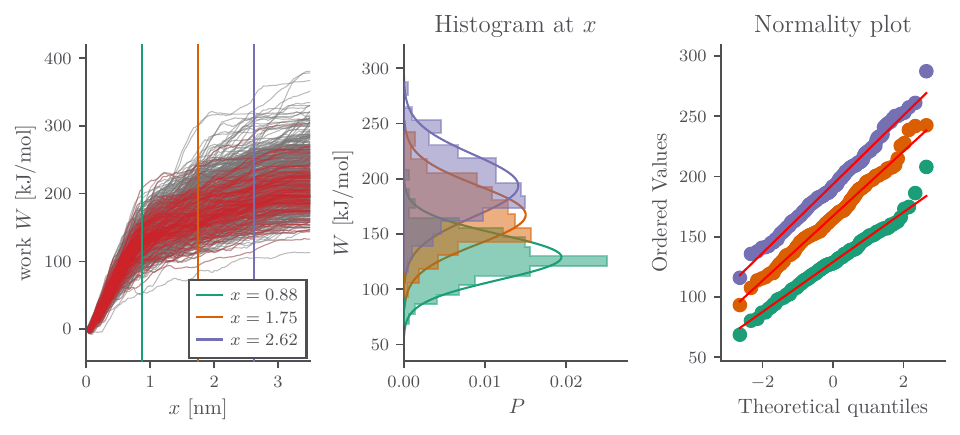}
        \caption{$d_\text{E}$ $c_2$ $\gamma=Q_3$}
    \end{subfigure}
    \begin{subfigure}{0.49\linewidth}
        \includegraphics[width=0.9\linewidth]{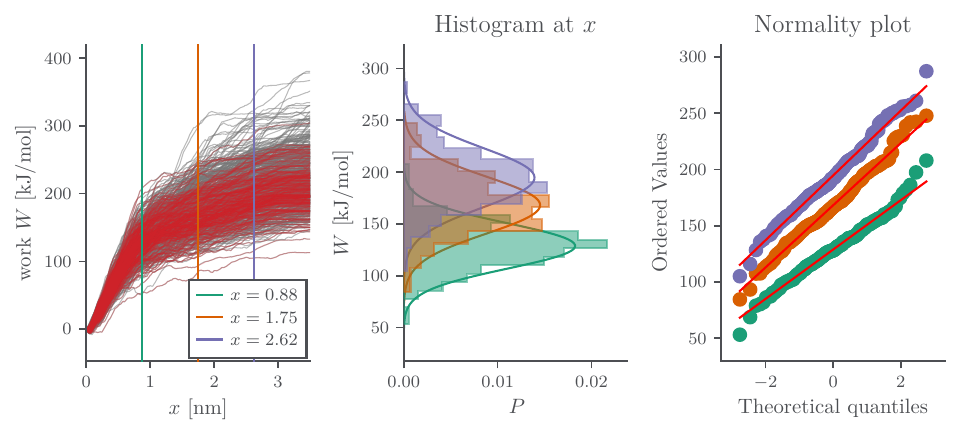}
        \caption{$d_\text{W}$ $c_1$ $\gamma=\langle Q\rangle_{23}$}
    \end{subfigure}
    \begin{subfigure}{0.49\linewidth}
        \includegraphics[width=0.9\linewidth]{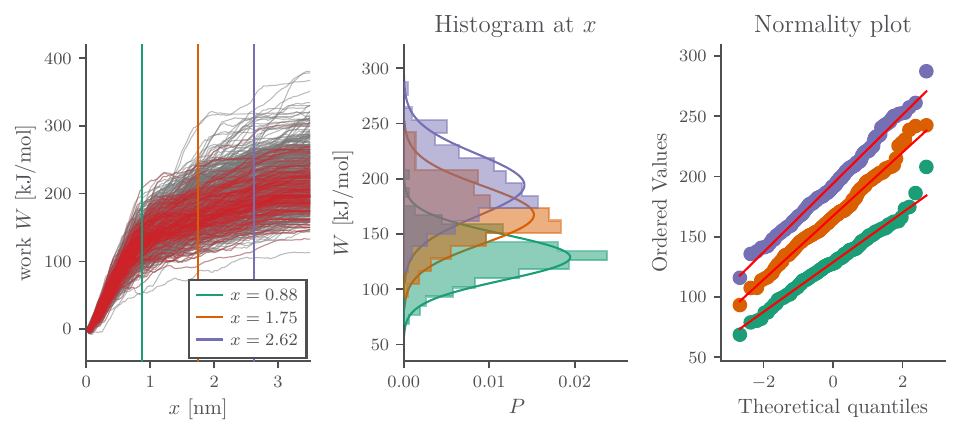}
        \caption{$d_\text{W}$ $c_1$ $\gamma=Q_3$}
    \end{subfigure}

    \caption{Work distribution analysis of clusters in Fig.~\ref{fig:SI_A2A_geometric_dG}. The similarity measure, cluster index and used $\gamma$ is indicated in the subcaptions. Left: Work values of all trajectories in gray, overlaid by the trajectories in the respective cluster in red. Middle: Histograms at values of $x$ indicated in the left plot. Right: Corresponding QQ plots.}
    \label{fig:SI_A2A_work_QQ1}
\end{figure}

\begin{figure}
    \centering
    \begin{subfigure}{0.49\linewidth}
        \includegraphics[width=0.9\linewidth]{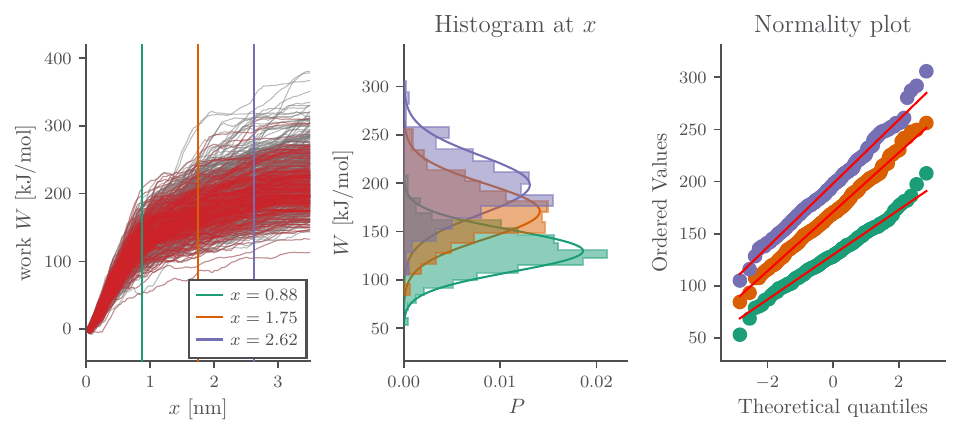}
        \caption{$d_\text{DTW}$ $c_1$ $\gamma=Q_2$}
    \end{subfigure}
    \begin{subfigure}{0.49\linewidth}
        \includegraphics[width=0.9\linewidth]{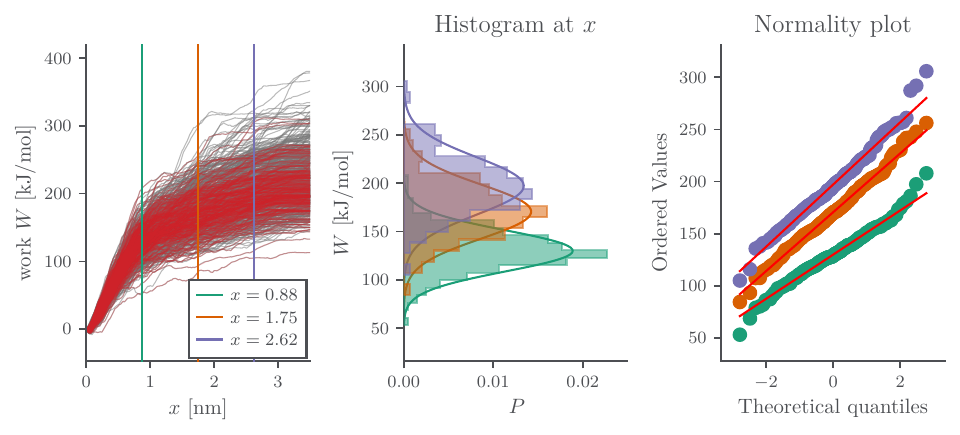}
        \caption{$d_\text{DTW}$ $c_1$ $\gamma=\langle Q\rangle_{23}$}
    \end{subfigure}
    \begin{subfigure}{0.49\linewidth}
        \includegraphics[width=0.9\linewidth]{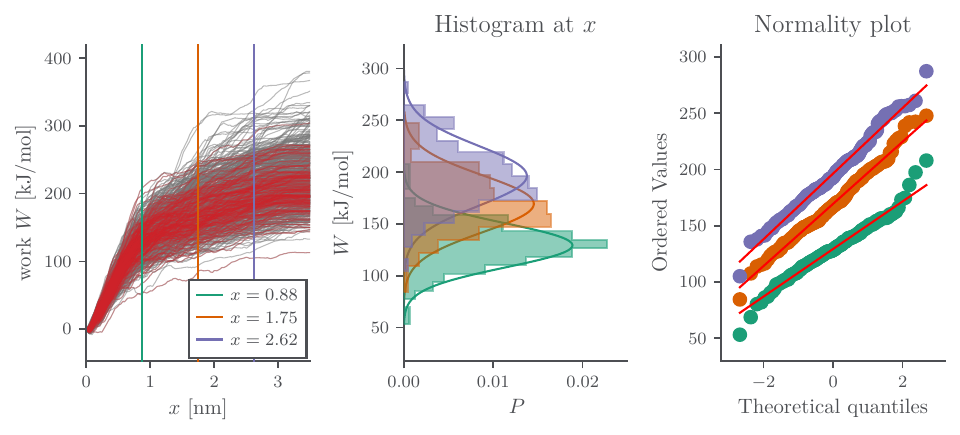}
        \caption{$d_\text{DTW}$ $c_2$ $\gamma=Q_3$}
    \end{subfigure}
    \begin{subfigure}{0.49\linewidth}
        \includegraphics[width=0.9\linewidth]{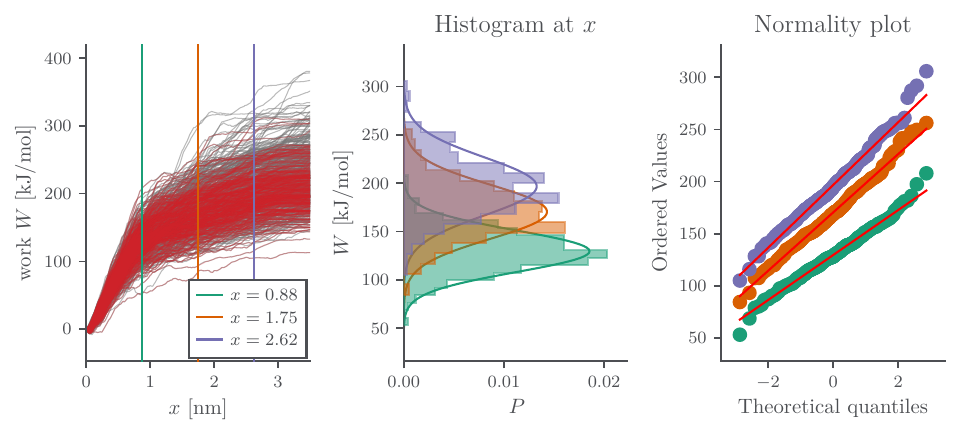}
        \caption{$d_\text{P}$ $c_1$ $\gamma=Q_2$}
    \end{subfigure}
    \begin{subfigure}{0.49\linewidth}
        \includegraphics[width=0.9\linewidth]{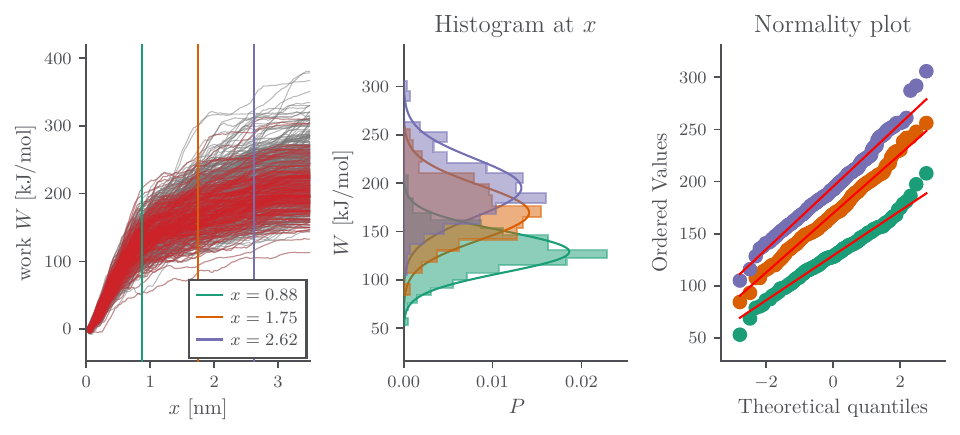}
        \caption{$d_\text{P}$ $c_1$ $\gamma=\langle Q\rangle_{23}$}
    \end{subfigure}
    \begin{subfigure}{0.49\linewidth}
        \includegraphics[width=0.9\linewidth]{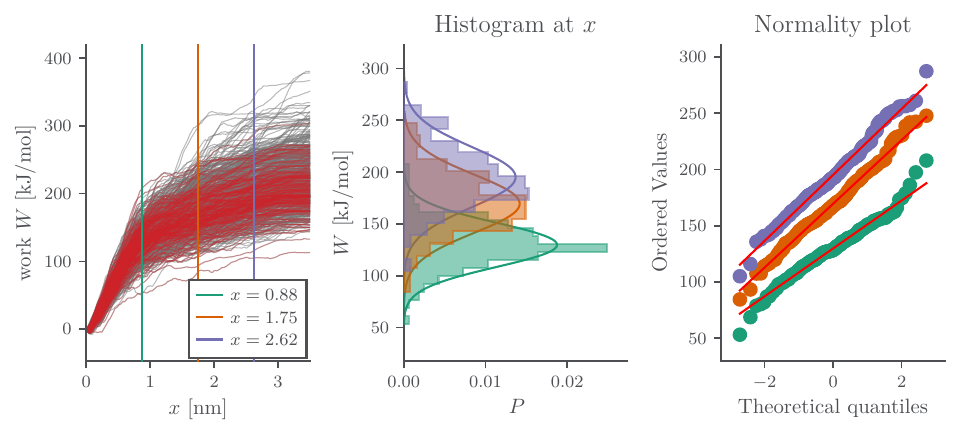}
        \caption{$d_\text{P}$ $c_1$ $\gamma=Q_3$}
    \end{subfigure}
    \caption{Work distribution analysis of clusters in Fig.~\ref{fig:SI_A2A_geometric_dG}. The similarity measure, cluster index and used $\gamma$ is indicated in the subcaptions. Left: Work values of all trajectories in gray, overlaid by the trajectories in the respective cluster in red. Middle: Histograms at values of $x$ indicated in the left plot. Right: Corresponding QQ plots.}
    \label{fig:SI_A2A_work_QQ2}
\end{figure}

\end{document}